\documentclass[%
 reprint,
nofootinbib,
 amsmath,amssymb,
 aps,
 prd,
]{revtex4-1}

\usepackage{graphicx}
\usepackage{dcolumn}
\usepackage{bm}
\usepackage{microtype}
\usepackage{hyperref}
\begin{document}


\title{Spherical collapse in coupled quintessence with a $\Lambda$CDM background}

\author{Bruno J. Barros$^{1}$, Tiago Barreiro$^{1,2}$ and Nelson J. Nunes$^{1}$}
\affiliation{$\,^{1}$Instituto de Astrof\'isica e Ci\^encias do Espa\c{c}o,\\ 
Faculdade de Ci\^encias da Universidade de Lisboa,  \\ Campo Grande, PT1749-016 
Lisboa, Portugal \\ $\,^{2}$Departamento de Matem\'atica,  ECEO, Universidade Lus\'ofona de Humanidades e 
Tecnologias, Campo Grande, 376,  1749-024 Lisboa, Portugal}

\date{\today}

\begin{abstract}
In this work we study the growth of cold dark matter density perturbations in the nonlinear regime on a conformally coupled quintessence model in which the background is designed to mimic a $\Lambda$CDM cosmology. The spherical collapse of overdense regions is analyzed. We highlight the role of the coupling on the overall dynamics, trace the evolution of the density contrast throughout the cosmic history and compute perturbative parameters such as the critical density contrast. We find that the coupling has the influence of delaying the collapse due to the slower growth of matter perturbations. We follow to compute the cluster number counts using the Press-Schechter and Sheth-Tormen mass functions. In both cases, the transfer of energy between the dark energy field and dark matter suppresses the number of objects at low redshifts and enhances the number at high redshifts. Finally, we compute the expected cluster number counts for the future eROSITA mission and the current South Pole Telescope survey.
\end{abstract}

\maketitle


\section{Introduction}

The mechanisms driving formation of structures in the Universe can properly be explored by tracking the evolution of matter overdensities in the nonlinear regime. Besides N-body simulations \cite{jennings2012,Maccio:2003yk,Baldi:2010vv,10.1111/j.1365-2966.2011.18894.x}, the spherical collapse model \cite{Gunn:1972sv,10.1111/j.1365-2966.2010.16841.x,Padmanabhan1999,Liddle:1993fq} has been proven to be a fruitful semianalytic method to explore the dynamics of these overdensities in the earliest stages of their nonlinear regime. Considering a spherical overdense patch in the Universe, we are able to witness the growth, and subsequently the collapse, of this region due to the gravitational pull attributed to the matter perturbation $\delta_m$ within it. If one considers models beyond the standard $\Lambda$CDM, there might be significant modifications on the dynamics governing the evolution of $\delta_m$. For example, the coupled quintessence models \cite{Wetterich:1994bg,Amendola:1999er,Barreiro:1999zs,Zimdahl:2001ar,Farrar:2003uw,Amendola:2007yx} where dark energy \cite{Tsujikawa:2010sc,Copeland:2006wr,detao} is described by a canonical scalar field $\phi$ allowed to interact with the matter species. Through this coupling, the scalar degree of freedom  mediates a fifth force \cite{Wintergerst:2010ui,Baldi:2010vv,Wintergerst:2009fh}, sourcing the gravitational potential through the Poisson equation and consequently  affecting the equation describing the dynamics of matter perturbations. There  might appear  additional effects, such as damping terms which  also suppress or enhance the growth of these fluctuations. Therefore, it is of great interest to study the influence of having such dark energy couplings on the evolution of overdensities and consequently on the formation of structure.

Spherical collapse in dynamical dark energy cosmologies has already been extensively studied in the literature \cite{Pace:2013pea,Delliou:2012ik,Nazari-Pooya:2016bra,Mota:2008ne}. The standard minimally coupled quintessence scenario analysis was conducted in \cite{Mota:2004pa} where the influence of assuming different types of scalar field potentials on the spherical collapse dynamics was studied. The scalar field perturbations on the collapse was also carefully explored, assuming that dark energy may cluster alongside with matter. It was shown in \cite{He:2010ta} that a collapse with a coupling within the dark sector can leave particular imprints on the cluster number density.
A thorough examination of the collapse for four different dark energy models was carried out in \cite{Nunes:2005fn}.  It was found that departures from the standard $\Lambda$CDM scenario may occur, and that these can be enhanced if there are inhomogeneities in the dark energy component. An analysis of interacting dark energy-dark matter cosmologies with a time varying coupling was explored in \cite{Baldi:2010vv}. The specific form of the coupling naturally has an impact in the background cosmology. Hence, together with N-body simulations, it was found that the formation of cosmic structure and the nonlinear demeanor of matter perturbations are strongly dependent on the background evolution.

Indeed, the $\Lambda$CDM model has endured throughout most observational tests hitherto apart from some underlying theoretical motivations regarding Einstein's cosmological constant \cite{Peebles:2002gy,Weinberg:2000yb,Zlatev:1998tr,Chimento2003,Zlatev1999,Cai:2004dk,Chimento:2003iea}. Hence, most models of dynamical dark energy proposed in the literature do not present large deviations from the standard model, particularly at the background level. Perturbatively however, there is a tension between redshift space distortions and the Planck data \cite{Macaulay2013,Battye:2014qga,Macaulay2013} in the amplitude of the matter power spectrum at the 8 Mpc scale, denoted by $\sigma_8$. In this regard, it was shown in \cite{Barros:2018efl}, that it is possible to construct a coupled quintessence model with the particularity of being able to mimic the exact same background as $\Lambda$CDM, but still being distinguishable at perturbative level. This is done by imposing {\it a posteriori} that the Hubble rate $H(z)$ matches the one of the standard $\Lambda$CDM model. In this way, background observations such as supernovae type Ia distances or baryonic acoustic oscillation observables, which are expressed only in terms of $H(z)$, cannot discriminate between the two models. In \cite{Asghari:2019qld} (see also \cite{Simpson:2010vh,Baldi:2016zom,Kumar:2017bpv}) it was presented an interacting dark energy scenario with the same behavior at the background level however in a different fashion: by considering couplings expressed in terms of the comoving 4-velocities of dark matter and dark energy the background cosmology is left unaffected, altering only the dynamics of inhomogeneous perturbations. There is at present a tension of $4.4\sigma$ on the background observable $H_0$ between the Cosmic Microwave Background measurements \cite{Aghanim:2018eyx} and the Cepheid variable-calibrated Type Ia supernovae \cite{Riess:2019cxk}. Since we are fixing our Hubble rate as $\Lambda$CDM, our model does not avoid this issue and we will not address it further in this paper.

This manuscript is organized as follows: Sec.~\ref{model} exposes the model that is adopted in this work, the background dynamics and the procedure in order to fix the background. In Sec.~\ref{sc} we discuss the spherical collapse model and present the nonlinear equations for the theory. We numerically solve the equations in Sec.~\ref{scI}, compute the value of the linear density contrast at collapse and analyze the solutions.  The halo number counts for the Press-Schechter and Sheth-Tormen mass functions are computed in Secs.~\ref{ps} and \ref{st}, encompassing a spherical and an ellipsoidal collapse, respectively. In Sec.~\ref{observations}, we estimate the cluster number counts that can be measured by the two surveys. Namely the  eROSITA satellite mission \footnote{https://www.mpe.mpg.de/eROSITA} \cite{Merloni:2012uf,Pillepich:2011zz} which was recently lauched, and  the South Pole Telescope\footnote{https://pole.uchicago.edu/} \cite{Ruhl:2004kv,deHaan:2016qvy,Bleem:2014iim}, in particular the SPT-SZ survey. We discuss the possibility of their ability to discriminate between models. Finally we conclude in Sec.~\ref{conclusions}.

\section{Model}
\label{model}

Our background cosmology will stand upon a flat Friedmann-Lema\^itre-Robertson-Walker (FLRW) line element 
\begin{equation}
\label{metric}
ds^2 = -dt^2 +a(t)^2 \delta_{ij}dx^idx^j,
\end{equation}
where $a(t)$ is the scale factor and $t$ the cosmic time.

In the present work, dark energy is described by a canonical scalar field $\phi$, the quintessence field \cite{Amendola:1999er,Wetterich:1994bg}, with energy density and pressure, respectively, given by,
\begin{eqnarray}
\rho_{\phi} &=& \frac{1}{2}\dot{\phi}^2 + V(\phi), \\
p_{\phi} &=& \frac{1}{2}\dot{\phi}^2 - V(\phi),
\end{eqnarray}
where $V(\phi)$ is the scalar potential.

We  assume that the quintessence field may couple to a pressureless cold dark matter (CDM) component with energy density $\rho_c$. This interaction within the dark sector can be expressed through the conservation relations, considering conformal couplings of the form \cite{Barros:2018efl,Barros:2019rdv,Amendola:2001rc,Amendola:2003wa,Amendola:2014kwa,Teixeira:2019tfi,Amendola:1999dr},
\begin{eqnarray}
\nabla_{\mu}T^{(c)}\,^{\mu}_{\nu} &=& \kappa \,\beta\rho_c\nabla_{\nu} \phi, \label{consDM}\\
\nabla_{\mu}T^{(\phi)}\,^{\mu}_{\nu} &=& -\kappa \,\beta\rho_c\nabla_{\nu} \phi, \label{consPhi}
\end{eqnarray}
where $\kappa^2=8\pi G$, $\nabla$ is the covariant derivative and $\beta$ is a constant expressing the strength of the coupling, governing the energy flow between the dark species. Thus, the individual energy-momentum tensors of dark energy and dark matter are not conserved, though the total energy-momentum tensor of the theory is. Regarding the action formalism for these coupled theories we refer the reader to \cite{Barros:2019rdv,PhysRevLett.64.123,Koivisto:2005nr,Bean:2000zm}.

We  also consider a noninteracting radiation component, consisting of photons and relativistic neutrinos, where the identity
\begin{equation}
\label{consRad}
\nabla_{\mu}T^{(r)}\,^{\mu}_{\nu} = 0
\end{equation}
holds.

In this setting, our species evolve in the FLRW background geometry as,
\begin{eqnarray}
\ddot{\phi} + 3H\dot{\phi} + V_{,\phi} &=& \kappa \beta\rho_c, \label{motion_phi} \\
\dot{\rho}_c + 3H\rho_c &=& -\kappa \beta \dot{\phi}\rho_c, \label{continuityc} \\
\dot{\rho}_r+4H\rho_r &=& 0,
\end{eqnarray}
where $V_{,\phi}$ is the scalar field potential derivative with respect to $\phi$ and $H=\dot{a}/a$, the Hubble rate. The Friedmann and Raychaudhuri equations read
\begin{eqnarray}
\frac{3}{\kappa^2}H^2 &=& \rho_c+\rho_{r}+\rho_{\phi}, \label{Friedmann} \\
-\frac{2}{\kappa^2}\dot{H} &=& \rho_c+\frac{4}{3}\rho_{r}+\dot{\phi}^2. \label{Raych}
\end{eqnarray}

We follow to fix the background to be the same as in the standard $\Lambda $CDM model, following the same procedure as in \cite{Barros:2018efl}. This can be achieved through the assumption that the Hubble rates coincide, i.e. $H(t)=H_s(t)$, where $H_s$ is the Hubble rate of the standard $\Lambda$CDM model
\begin{equation}
\label{friedmann}
\frac{3}{\kappa^2}H_s^2 = \rho_{cdm}+\rho_{r} + \rho_{\Lambda},
\end{equation}
where $\rho_{\Lambda}$  and $\rho_{cdm}=\rho_{cdm}^0\,a^{-3}$ are the energy densities of the cosmological constant and standard cold dark matter, respectively. Note that we only assume that the Hubble rates are the same, not the individual evolution for each species. With this assumption, we are able to find the particular form of the potential function that guarantees the condition $H=H_s$. Hence, $V$ is obliged to follow the dynamics of
\begin{equation}
\label{PotFixed}
V= \frac{1}{2}\dot{\phi}^2 + \rho_{\Lambda}.
\end{equation}
Thus, the respective energy densities for the quintessence and the coupled cold dark matter components can be written, respectively, as:
\begin{eqnarray}
\rho_{\phi} &=& \dot{\phi}^2 + \rho_{\Lambda}, \\
\rho_c &=& \rho_{cdm} - \dot{\phi}^2. \label{rholcdm}
\end{eqnarray}
We refer the reader to \cite{Barros:2018efl} for details. Taking the derivative with respect to (wrt) $\phi$ of Eq.~\eqref{PotFixed} and substituting in Eq.~\eqref{motion_phi} we find the background equation of motion for the scalar field which renders the background to a $\Lambda$CDM evolution,
\begin{equation}
\label{motion_phi_fixed}
2\ddot{\phi}+\dot{\phi} \,( 3H-\kappa\beta\dot{\phi} ) - \kappa\beta\rho_{cdm} =0.
\end{equation}
We can write Eq.~\eqref{motion_phi_fixed} using derivatives wrt the number of e-folds $N=\ln a$, i.e. $\phi':=\partial \phi/\partial N = \dot{\phi}/H$, as
\begin{equation}
\label{motionPhi}
2\phi''+\phi'\left( 3 + 2\frac{H'}{H} + \kappa\beta\phi'\right)-\frac{3}{\kappa} \beta\,\Omega_{cdm} =0,
\end{equation}
where
\begin{equation}
\label{H}
\frac{H'}{H} = -\frac{1}{2}\left( 3+\Omega_r - 3\,\Omega_{\Lambda} \right)
\end{equation}
and we have introduced the relative energy density parameter of the $i^{\rm th}$-species,
\begin{equation}
\Omega_i = \frac{\kappa^2}{3}\frac{\rho_i}{H^2}.
\end{equation}

The study of background and first order perturbations in this present model were conducted in \cite{Barros:2018efl}. The linear evolution of density perturbations is of great interest in cosmology, as it can provide direct observables, such as the power spectrum and the $\sigma_8$ parameter, which can be directly linked to observations. Nonetheless, there are certain phenomena that can only be captured by studying the nonlinear regime (see Sec.~8 of \cite{Liddle:1993fq}). In the following section, we investigate some of these phenomena, in particular, the spherical collapse of matter fluctuations and the number of bound objects formed with a certain mass range at a given redshift \cite{Liddle:1995ay}, and investigate the influence of the coupling on this quantity.

\section{Spherical Collapse}
\label{sc}

Let us consider a CDM density perturbation, $\delta = \delta\rho_c/\bar{\rho}_c \ll 1$ (a bar denotes background quantities). As the perturbation grows along with its expanding background, at some point, depending on its scale, it may grow close to unity where the linear regime breaks down. Therefore, in order to have a grasp on the mechanisms driving structure formation we need to understand the nonlinear regime \cite{Amendola:2003wa,LeDelliou:2005ig}.

An overdense region of radius $r$ first grows in size with the Hubble expansion but sooner or later, depending on the scale, it departs from the latter and collapses. The {\it spherical collapse model} \cite{Gunn:1972sv,Liddle:1993fq,Nunes:2004wn} is an approach to trace the evolution of the perturbations on the primary phases of their nonlinear regime. This procedure assumes a certain overdense spherical (and nonrotating) region with a certain radius $r(t)$. Birkhoff's theorem \cite{1923rmpbookB} states that the evolution of this radius depends solely on its enclosed mass. Hence, we can model this region as a subuniverse with $\rho_c = \bar{\rho}_c + \delta \rho_c$ with ``scale factor'' $r$,
\begin{equation}
\left( \frac{\dot{r}}{r} \right)^2 = \frac{\kappa^2}{3}\sum_i \rho_i -\frac{K}{r^2},
\end{equation}
where the sum is over all the $i$-th species.
The presence of a curvature term simply manifests that the spherical patch is positively curved as its density is larger than its critical (background) one due to the presence of the overdensity $\delta \rho_c$ \cite{Wintergerst:2010ui}. Note that the background quantity $\bar{\rho}_c$ evolves according to the standard Friedmann Eq.~\eqref{friedmann},
\begin{equation}
\left( \frac{\dot{a}}{a} \right)^2 = \frac{\kappa^2}{3}\sum_i \bar{\rho}_i .
\end{equation}

The main assumption of the spherical collapse model is that the overdensity $\delta$ follows a top hat (or step) function \cite{detao}, where
\begin{equation}
\delta = \frac{\rho_c}{\bar{\rho}_c} -1
\end{equation}
inside the spherical region, and $\delta = 0$ outside.

Assuming that initially the scale factors $r$ and $a$ are equal, i.e. $r_{in} = a_{in}$, and that the mass of the CDM particles of the background are the same as on the spherical overdense region, we may write \cite{Wintergerst:2010ui}
\begin{equation}
\label{delta_r}
1+\delta = (1+\delta_{in})\left( \frac{a}{r} \right)^3,
\end{equation}
where $\delta_{in}=\delta(z_{in})$ is the initial density contrast for the cold dark matter component.
From Eq.~\eqref{delta_r} it becomes evident that the divergence of the density contrast, $\delta\rightarrow\infty$, happens as the region collapses, $r\rightarrow 0$.

The second order equation for the evolution of the perturbations in coupled quintessence, in the small scales regime (Newtonian limit), were derived in \cite{Wintergerst:2010ui} (see also \cite{Pace:2013pea,Savastano:2019zpr}) from the full set of nonlinear hydrodynamical equations, and reads,
\begin{eqnarray}
\label{nleq1}
\ddot{\delta} &+& \dot{\delta} \left( 2H - \kappa \beta \dot{\phi}\right) - \frac{\kappa^2}{2} \bar{\rho}_c\,\delta \left( 1+\delta \right)\left( 1+2\beta^2 \right)  \nonumber \\
&-& \frac{4}{3}\frac{\dot{\delta}^2}{1+\delta} = 0.
\end{eqnarray}
The presence of the scalar field results in the emergence of a fifth force, where the CDM component  experiences an effective gravitational constant $G_{\rm eff} = G_N (1+2\beta^2)$ \cite{Wintergerst:2010ui,Amendola:2012ys}. It also adds an extra contribution to the frictional term proportional to $\beta \dot{\phi}$, which in our case always weakens the overall damping effect, since $\beta\dot{\phi}>0$ as discussed in \cite{Barros:2018efl}. The balance of these two effects has a direct impact on the growth rate of the matter perturbations \cite{Pettorino:2008ez,Leithes:2016xyh}. In the present work, the background is fixed in order to reproduce $\Lambda$CDM. Hence, we may write Eq.~\eqref{nleq1} replacing $\bar{\rho}_c$ using Eq.~\eqref{rholcdm},
\begin{eqnarray}
\label{nleq2}
\ddot{\delta} &+& \dot{\delta} \left( 2H - \kappa \beta \dot{\phi}\right) - \frac{\kappa^2}{2} \left(\bar{\rho}_{cdm} - \dot{\phi}^2\right)\,\delta \left( 1+\delta \right)\left( 1+2\beta^2 \right)  \nonumber \\
&-& \frac{4}{3}\frac{\dot{\delta}^2}{1+\delta} = 0,
\end{eqnarray}
which can be written with derivatives with respect to $N$ as
\begin{eqnarray}
\delta'' &+& \delta' \left( 2+ \frac{H'}{H} - \kappa \beta \phi'\right) - \frac{3}{2} \left(\Omega_{cdm} - \frac{\kappa^2}{3}\phi'^2\right)  \nonumber \\
&\times &  \delta \left( 1+\delta \right)\left( 1+2\beta^2 \right) - \frac{4}{3}\frac{\delta'^2}{1+\delta} = 0, \label{nleq3}
\end{eqnarray}
where $H'/H$ is given by Eq.~\eqref{H}. Linearizing Eq.~\eqref{nleq3} we recover the first order equation studied in \cite{Barros:2018efl},
\begin{eqnarray}
\label{nleq4}
\delta'' &+& \delta' \left( 2+ \frac{H'}{H} - \kappa \beta \phi'\right) - \frac{3}{2} \left(\Omega_{cdm} - \frac{\kappa^2}{3}\phi'^2\right)  \nonumber \\
&\times &  \delta \left( 1+2\beta^2 \right) = 0.
\end{eqnarray}

\begin{figure}[t]
\begin{center}
\includegraphics[width=0.48\textwidth]{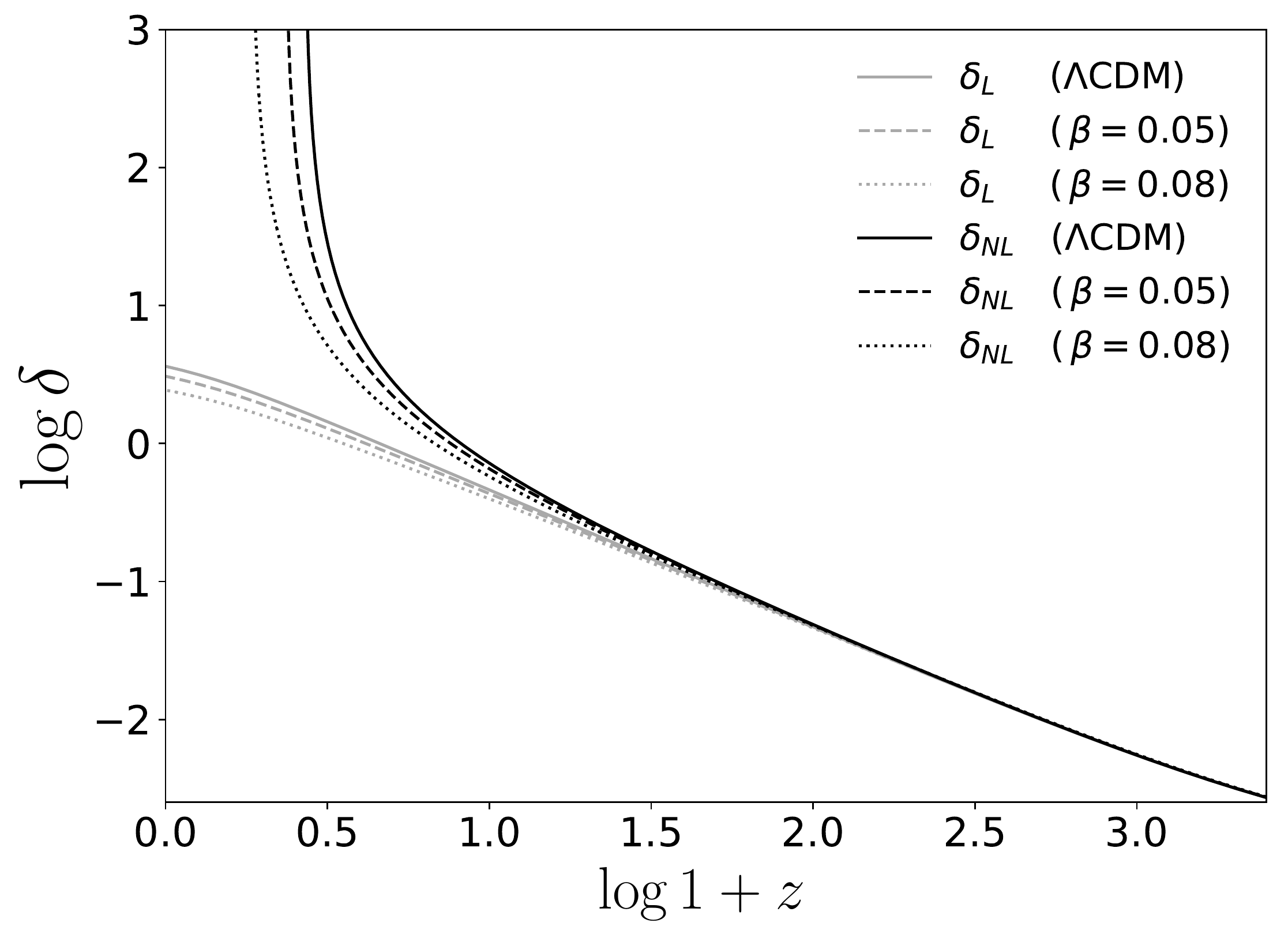}
\end{center}
\caption{\label{deltas} Linear (gray), $\delta_L$, and nonlinear (black), $\delta_{NL}$, CDM density contrast versus redshift. Solutions of Eqs.~\eqref{motionPhi}, \eqref{nleq3} and \eqref{nleq4} for $\beta=0$ (solid), $\beta=0.05$ (dashed) and $\beta=0.08$ (dotted) with $\delta_{in} = 9\times 10^{-4}$.}
\end{figure}

In the following section we numerically evolve Eqs.~\eqref{nleq3} and \eqref{nleq4} together with the background Eq.~\eqref{motionPhi} and study how the spherical collapse parameters behave when the coupling parameter $\beta$ changes.

\subsection{The collapse of the matter density contrast}
\label{scI}

Our simulations start in a radiation domination era, at $N_{in}=-14$ ($z_{in} \approx 10^6$). The initial conditions for the quintessence field are taken to be $\phi_{in} = \dot{\phi}_{in}=0$, to guarantee that at early times the energy densities for the individual species coincide with $\Lambda$CDM (see \cite{Barros:2018efl} for details). Regarding the density contrast, we take $\dot{\delta}_{in}=0$ and $\delta_{in} < 10^{-3}$, well within the validity of the linear regime at early times \cite{Wintergerst:2010ui}. We fix the parameters using the latest Planck 2018 values \cite{Aghanim:2018eyx}, $\Omega^0_{cdm} = 0.311$, $\Omega^0_r h^2 = 4.1\times 10^{-4}$ and $\Omega_{\Lambda} = 1 - \Omega_c - \Omega_r$, and consider $\beta \geqslant 0$.

In \cite{Barros:2018efl} it was found that in order to mimic the same background as $\Lambda$CDM, the amount of CDM today must decrease with increasing $\beta$. Consequently, this  leads to a slower growth of the matter fluctuations. This effect can be observed in Fig.~\ref{deltas} (gray lines). This is ascribed to the term multiplying $\delta$ in Eq.~\eqref{nleq4}  as it becomes smaller than $\Omega_{cdm}$ due to the presence of the kinetic term $-\kappa^2\phi'^2/3$ \cite{Barros:2018efl}.

\begin{figure}[t]
\begin{center}
\includegraphics[width=0.48\textwidth]{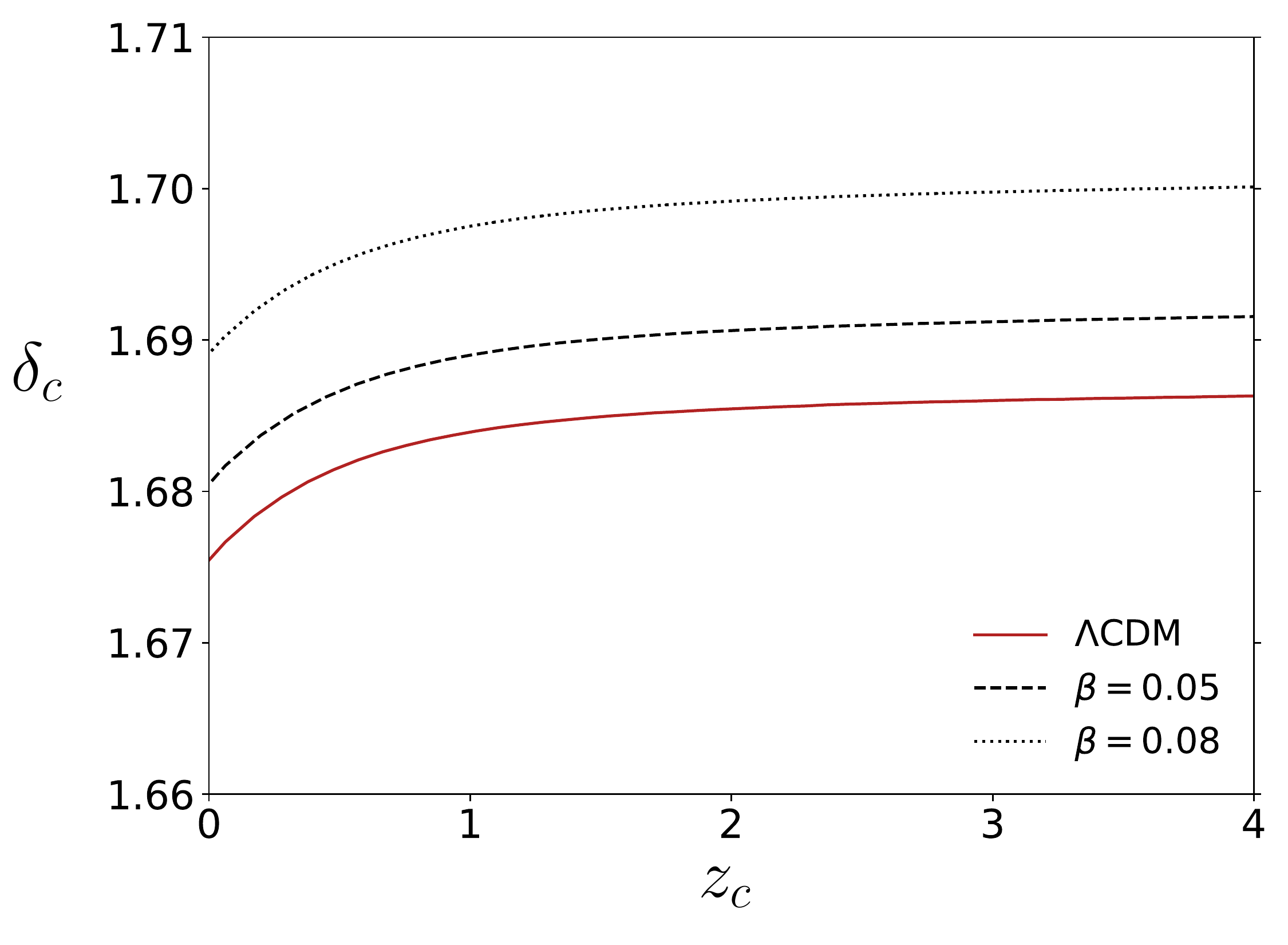}
\end{center}
\caption{\label{deltacrit} Linear density contrast at collapse, $\delta_c$, versus redshift of the collapse, $z_c$, for $\beta=0$ (solid), $\beta=0.05$ (dashed) and $\beta=0.08$ (dotted).}
\end{figure}

\begin{figure*}[t]
\begin{center}
\includegraphics[width=0.9\textwidth]{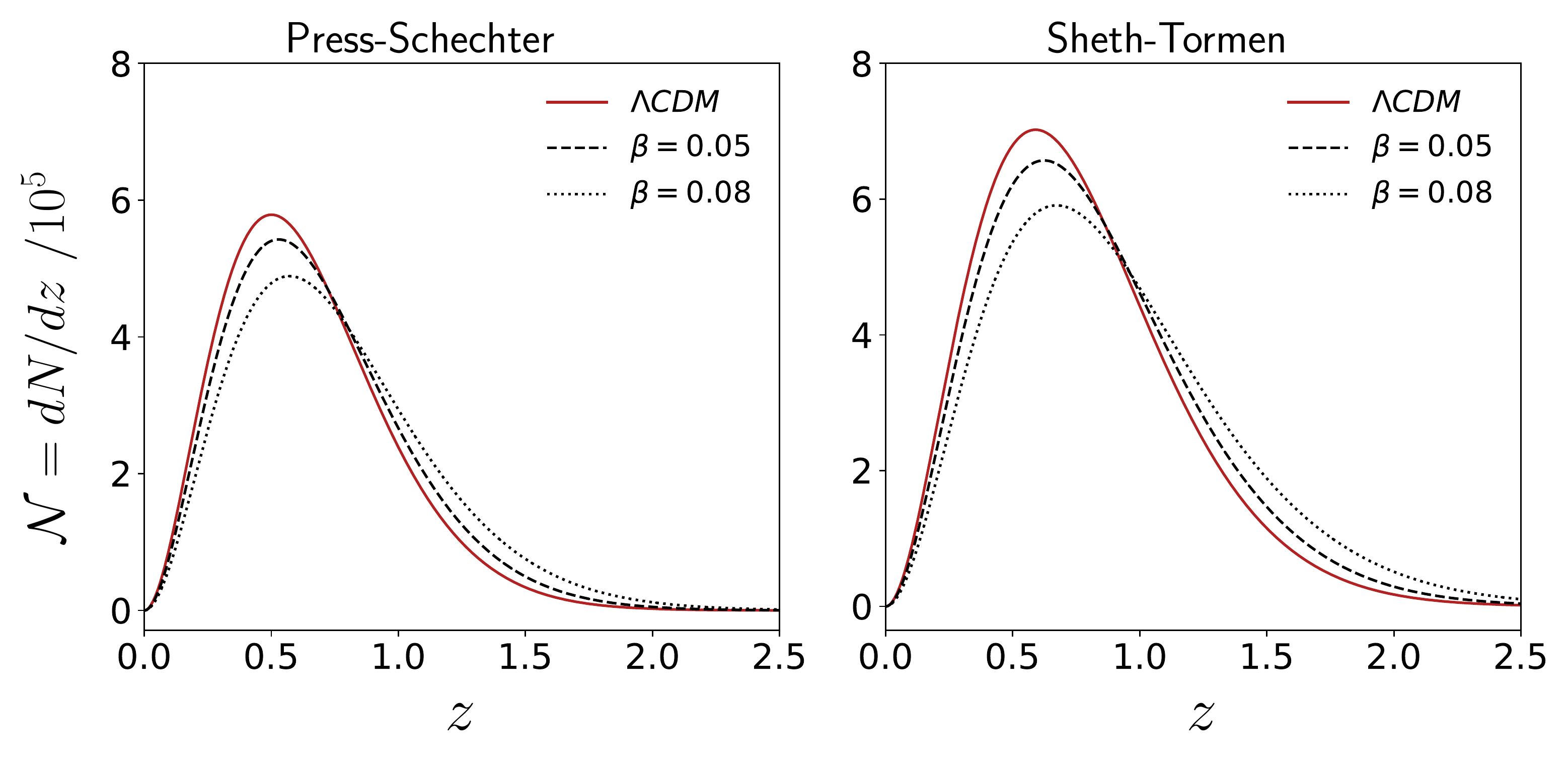}
\end{center}
\caption{\label{numberf} Comoving number of dark matter halos with masses within $10^{14}h^{-1}M_{\odot}<M<10^{16}h^{-1}M_{\odot}$ for the Press-Schechter (left panel) and the Sheth-Tormen (right panel) mass functions, with $\beta=0$ (solid), $\beta=0.05$ (dashed) and $\beta=0.08$ (dotted).}
\end{figure*}

The nonsolid lines of Fig.~\ref{deltas} portray the evolution of the nonlinear density contrast for different coupling values. As the perturbation grows, eventually the linear regime is broken, and the nonlinear terms start to dominate the evolution.
When a sufficient amount of density contrast is gathered the collapse occurs, $\delta\rightarrow \infty$. As the growth of the matter fluctuation is slower for higher values of the coupling $\beta$, the perturbation  takes longer to assemble the critical amount of matter for the collapse to happen. Hence, as we observe in Fig.~\ref{deltas}, the collapse befalls latter  for larger values of $\beta$.

Another quantity that is useful to characterize the spherical collapse model is the critical density contrast. This is defined as the value of the linear density contrast $\delta_L$ when
the nonlinear density contrast diverges, i.e. when $\delta_{NL}\rightarrow\infty$. Running the simulation until $\delta_{NL}$ diverges, we can extract the value of the redshift of the collapse $z_c$ as well as the linear density contrast $\delta_c := \delta_L (z=z_c)$. 
We can then change the collapse time by
varying the initial condition $\delta_{in}$, therefore obtaining different values for $z_c$ and the corresponding $\delta_c$.
The results are shown in Fig.~\ref{deltacrit}, where $\Lambda$CDM ($\beta=0$) is plotted in solid as a reference. As expected, we observe that increasing the value for the coupling parameter leads to higher values of $\delta_c$. As the growth is slower (for higher values of $\beta$), a greater amount of density contrast is required for the collapse to happen. An opposite effect was found in \cite{Sapa:2018jja} considering disformal couplings \cite{Bekenstein:1992pj,Zumalacarregui:2013pma}.

\section{Press-Schechter formalism}
\label{ps}

One of the parameters computed in the last section was the critical density contrast $\delta_c$. This object is of great interest since it enters directly in the {\it Press-Schechter} formula \cite{Press:1973iz}, which allows us to calculate the number density of collapsed objects, in a given mass range, over a volume and at a specific time in the cosmic history. This formalism stands upon the assumption that the matter density field follows a Gaussian distribution \cite{Viana:1995yv}. The prediction for the comoving number density of collapsed objects with mass between $M$ and $M+dM$ is \cite{Manera:2005ct,PhysRevD.85.023503,Viana:1995yv}
\begin{eqnarray}
\frac{dn}{dM} = && -\sqrt{\frac{2}{\pi}}\frac{\tilde{\rho}_c(z)}{M}\frac{\delta_c(z)}{\sigma(z,M)}\frac{d\ln \sigma (z,M)}{dM} \nonumber \\
&&\times\exp\left[ -\frac{\delta_c(z)^2}{2\sigma(z,M)^2} \right], \label{press}
\end{eqnarray}
where $\tilde{\rho}_c := a^3\rho_c$ is the comoving matter density  \cite{Manera:2005ct,Liddle:1995ay} and the variance $\sigma(z,M)$ corresponds to the rms density fluctuation in a sphere of radius $R$, enclosing a mass $M$. We can express the variance in terms of the growth factor $g(z):=\delta(z)/\delta(0)$, at the fixed scale of $R=R_8 = 8h^{-1}$Mpc \cite{Nunes:2005fn}, as
\begin{equation}
\sigma (z,M) = \sigma(0,M_8)\left( \frac{M}{M_8} \right)^{-\gamma/3} g(z),
\end{equation}
where $M_8 = 6\times 10^{14}\Omega_c h^{-1}M_{\odot}$ is the mass within the sphere and, following \cite{Nunes:2005fn},
\begin{equation}
\gamma = \left( 0.3\Gamma + 0.2 \right)\left[ 2.92 + \frac{1}{3}\log\left( \frac{M}{M_8} \right) \right].
\end{equation}
In the present work, we use $\Gamma = \Omega_ch$ \cite{Sapa:2018jja} and $\sigma_8:= \sigma(0,M_8) = 0.811$ \cite{Aghanim:2018eyx}. We can convert the number density Eq.~\eqref{press} into the effective number of objects with masses between $M_{\rm inf}<M<M_{\rm sup}$ per redshift and square degree,
\begin{equation}
\label{number}
\mathcal{N} := \frac{dN}{dz} = \int_{1{\rm deg}^2}d\Omega \frac{dV}{dz\,d\Omega}\int_{M_{\rm inf}}^{M_{\rm sup}} \frac{dn}{dM}dM,
\end{equation}
where
\begin{equation}
\label{volume}
\frac{dV}{dz\,d\Omega} = \frac{c\,r(z)^2}{H(z)} = \frac{c}{H(z)}\left[ \int_0^z \frac{c}{H(x)}dx \right]^2
\end{equation}
is the comoving volume element, $r(z)$ being the comoving distance. One interesting feature observed in the present model, is that this object, Eq.~\eqref{volume}, is independent of the value of $\beta$, in contrast with the standard dynamical dark energy models \cite{Nunes:2005fn,Manera:2005ct,Sapa:2018jja,Pace:2010sn}. This is due to the fact that it depends solely on the Hubble rate $H(z)$ which we have assumed to always match the $\Lambda$CDM evolution.

\begin{figure}[t]
\begin{center}
\includegraphics[width=0.49\textwidth]{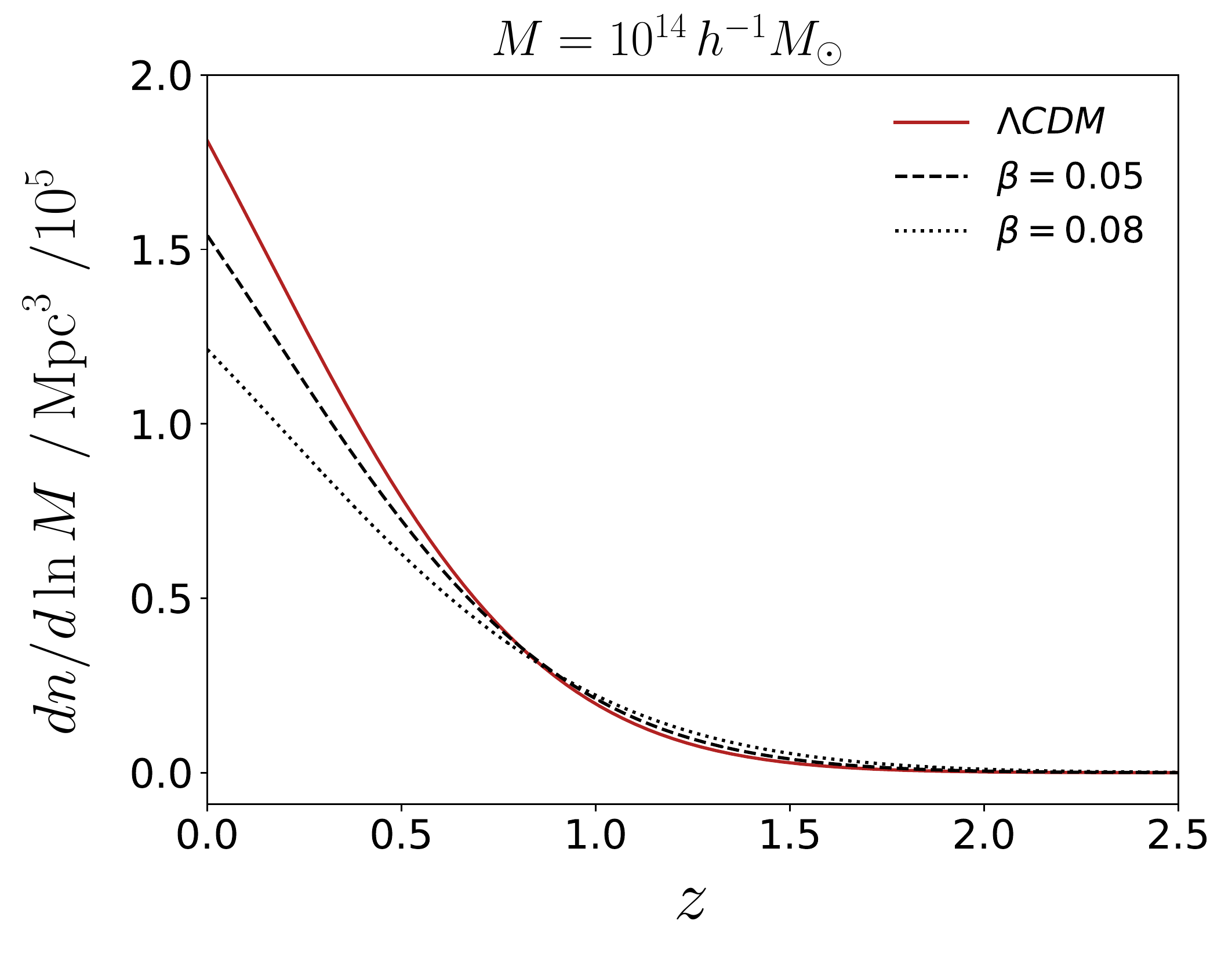}
\end{center}
\caption{\label{mass} Press-Schechter mass function, Eq.~\eqref{press}, for $M=10^{14}\,h^{-1}M_{\odot}$ with $\beta=0$ (solid), $\beta=0.05$ (dashed) and $\beta=0.08$ (dotted).}
\end{figure}

Our main goal in this section is to investigate the influence of the coupling on the number of dark matter halos formed. To this end we consider masses within the range of galaxy clusters, $10^{14}\,h^{-1}M_{\odot}<M<10^{16}\,h^{-1}M_{\odot}$ \cite{Nunes:2005fn}. In Fig.~\ref{numberf} we show the results for the comoving number counts of DM halos for $\beta = 0$, $\beta = 0.05$ and $\beta = 0.08$. We observe that the number counts are suppressed at low redshifts by the coupling and enhanced at high redshifts. This can be understood through two competing effects. On the one hand, the
$\Lambda$CDM model, having higher values of $\delta_c/\sigma$ compared to the coupled models, leads to smaller values of the mass function Eq.~\eqref{press} through the exponential term. On the other hand,  the background matter energy density $\tilde{\rho}_c$ is also higher for $\Lambda$CDM, leading to an increase in the mass function. This latter effect is dominant at lower redshifts, whereas the first dominates at higher redshifts causing a crossover between the curves of the expected number of clusters $\mathcal{N}$  as seen in Fig.~\ref{mass}.
 A similar behavior can be seen  in \cite{Nunes:2005fn}, however due to a completely different cause: the suppression of the Press-Schechter function at low redshifts was induced by deviations in the volume element when varying the equation of state parameter for the dark energy fluid.

\begin{figure}[t]
\begin{center}
\includegraphics[width=0.49\textwidth]{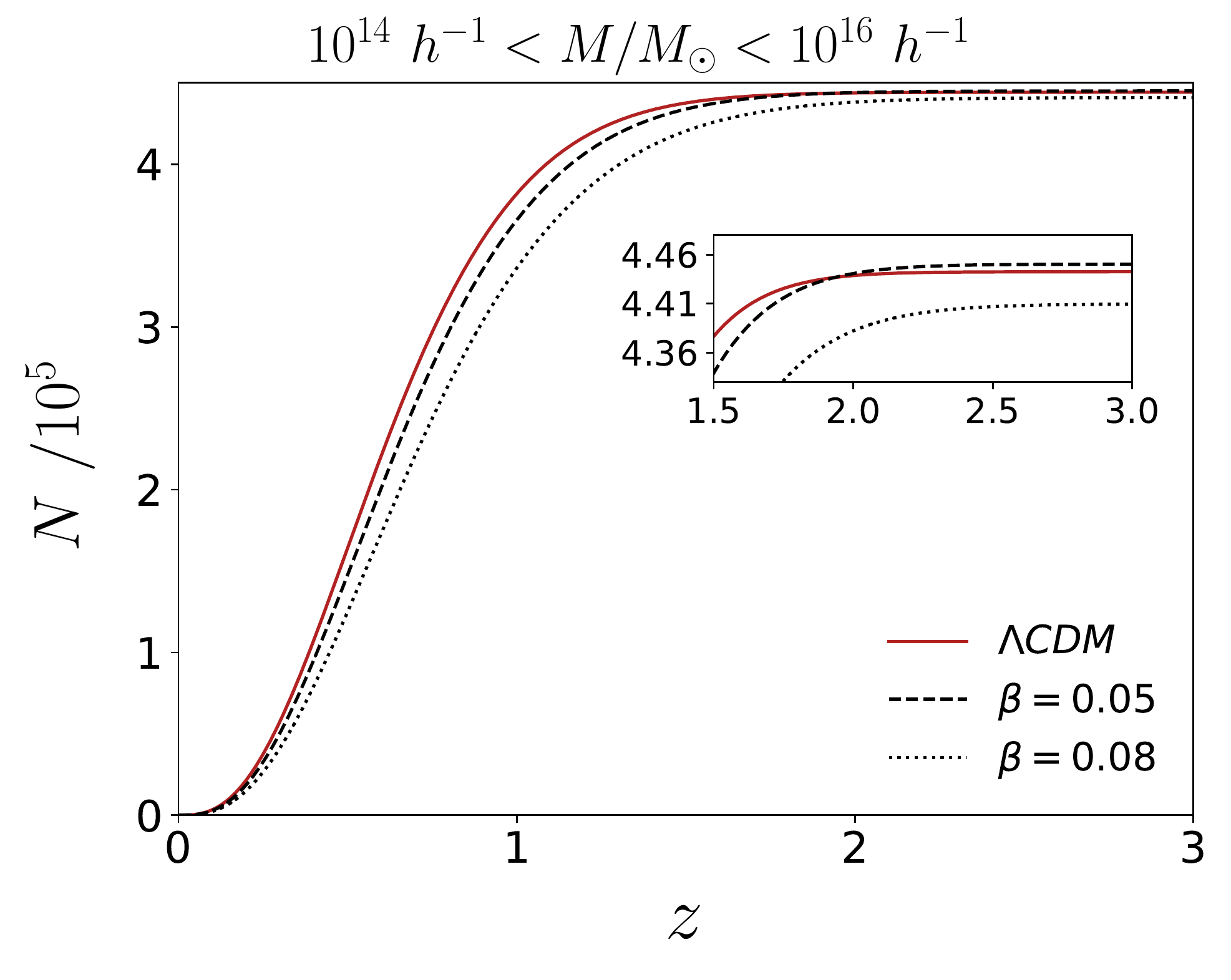}
\end{center}
\caption{\label{intN} Integrated number of dark matter halos with masses within $10^{14}h^{-1}M_{\odot}<M<10^{16}h^{-1}M_{\odot}$. Solution of Eq.~\eqref{Intnumber} with $\beta=0$ (solid), $\beta=0.05$ (dashed) and $\beta=0.08$ (dotted).}
\end{figure}

It is also useful to calculate the integrated number of objects in the full sky up to redshift $z$, simply by integration of Eq.~\eqref{numberf}, that is,
\begin{equation}
\label{Intnumber}
N = \int_{1{\rm deg}^2}d\Omega \int_{M_{\rm inf}}^{M_{\rm sup}} \int_0^z\frac{dn}{dM}\frac{dV}{d\bar{z}\,d\Omega}dMd\bar{z}.
\end{equation}
The results for $N$ are presented in Fig.~\ref{intN}. We observe that a coupling of $\beta = 0.05$ leads to a higher integrated number of dark matter halos at higher redshifts. On the other hand, for larger values of $\beta$ the suppression of the mass function at low $z$ is significantly more pronounced, ultimately causing the integrated number of halos to remain below $\Lambda$CDM even for higher redshifts.

\begin{figure}[t]
\begin{center}
\includegraphics[width=0.47\textwidth]{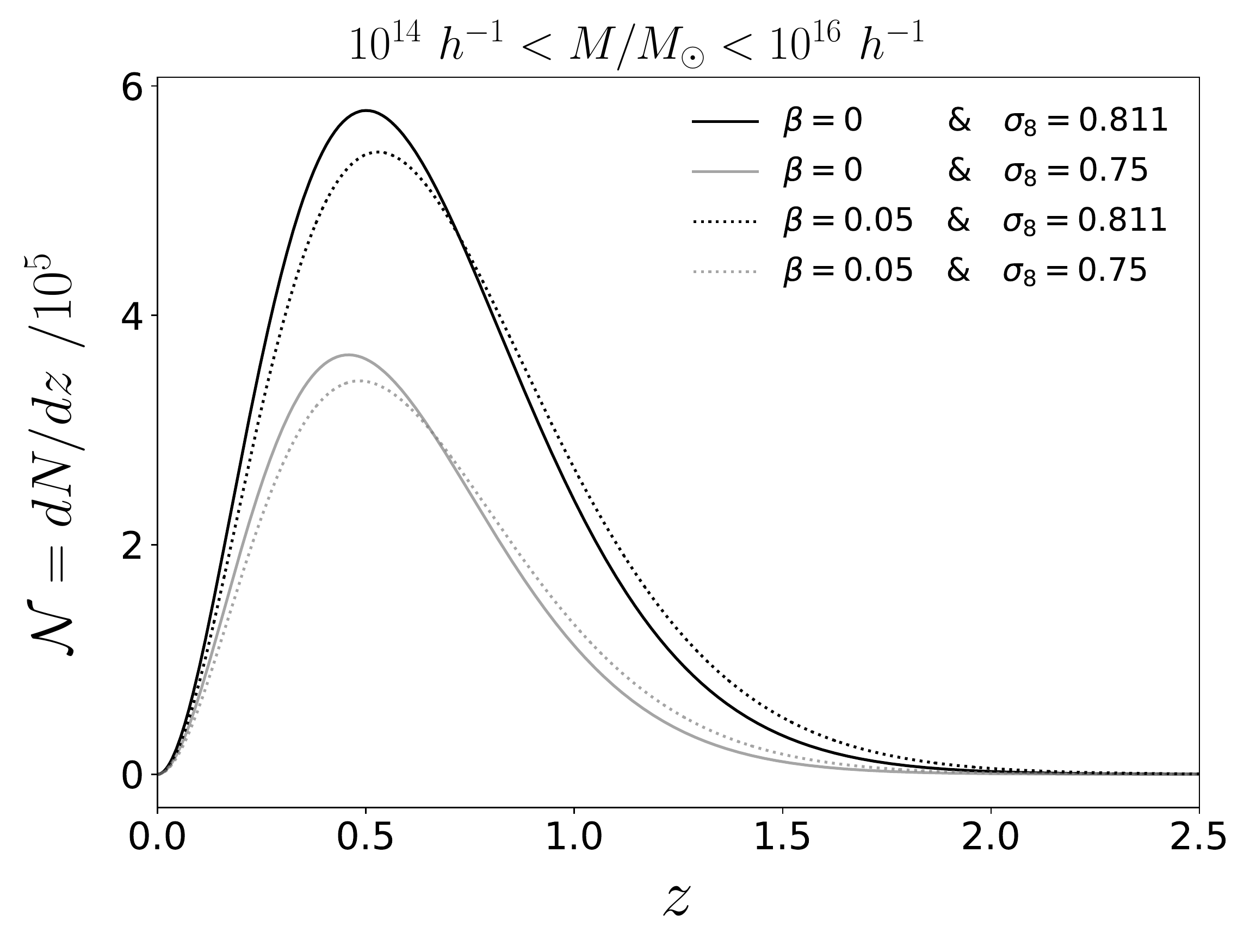}
\end{center}
\caption{\label{sigmaPlot} Comoving number of dark matter halos with masses within $10^{14}h^{-1}M_{\odot}<M<10^{16}h^{-1}M_{\odot}$ using the Press-Schechter mass function for the uncoupled case (solid) and $\beta=0.05$ (dotted) with $\sigma_8 = 0.811$ (black) and $\sigma_8 = 0.75$ (gray).}
\end{figure}

Since our aim was to focus solely on the role of the interaction, throughout this work we have assumed all parameters fixed, except the coupling $\beta$. 
However, the expected number counts are also influenced by other cosmological parameters, and in particular have a strong dependence on the value of $\sigma_8$. Increasing $\sigma_8$ significantly increases  the expected number counts. We illustrate this trend in Fig.~\ref{sigmaPlot}. Ultimately one needs to constrain the $\beta$ and $\sigma_8$ parameters simultaneously and in general we expect these to be correlated.

\section{Sheth-Tormen mass function}
\label{st}

The pioneering model proposed by Press and Schechter is successful in capturing a general picture of the distribution of objects in the Universe. Nonetheless, from simulations it is known \cite{Devi:2011gb} that the PS formula predicts a higher abundance of dark matter halos at low redshifts and a lower abundance at high $z$. This fact led Sheth and Tormen \cite{Sheth:1999mn} to formulate a modification of the PS formalism, assuming an ellipsoidal model for the collapse of the density contrast region, providing a modified mass function which seems to be in better agreement with simulations \cite{Reed:2006rw}. In this section we compare the numerical results for the abundance of dark matter halos from the last section, with the ones using the Sheth-Tormen formalism.

The Sheth-Tormen (ST) mass function can be written as
\begin{eqnarray}
\frac{dn}{dM} = && -A\sqrt{\frac{2a}{\pi}}\frac{\tilde{\rho}_c(z)}{M}\left[ 1+\left( \frac{\sigma(z,M)^2}{\delta_c(z)^2} \right)^p \right]\frac{\delta_c(z)}{\sigma(z,M)} \nonumber \\
&&\times\frac{d\ln \sigma (z,M)}{dM} \exp\left[ -\frac{a\delta_c(z)^2}{2\sigma(z,M)^2} \right], \label{stmass}
\end{eqnarray}
where $a$ and $p$ are parameters fitted by numerical simulations and $A$ is a normalization constant -- an assumption such that all of dark matter reside in halos -- (see, for example, \cite{Sheth:1999mn,Devi:2011gb} for details). Following \cite{Devi:2011gb,Devi:2014rva}, we use the values $(a,p,A)=(0.707,0.3,0.322)$ which are well in agreement with numerical simulations. Note that the standard Press-Schechter mass function is recovered for $(a,p,A)=(1,0,1/2)$.

\begin{figure}[t]
\begin{center}
\includegraphics[width=0.49\textwidth]{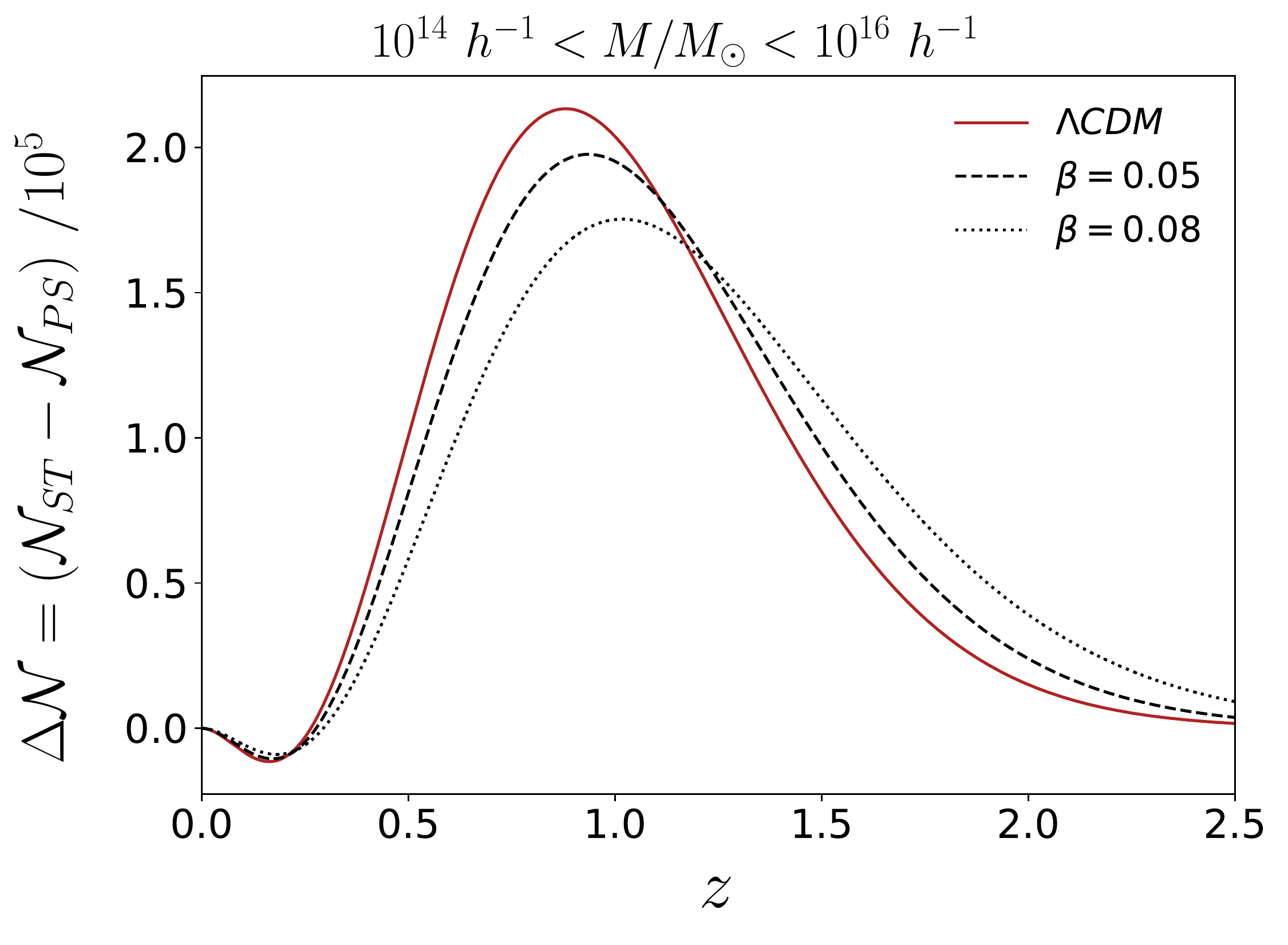}
\end{center}
\caption{\label{stfig2} Difference of the comoving number of dark matter halos with masses within $10^{14}h^{-1}M_{\odot}<M<10^{16}h^{-1}M_{\odot}$, between the ST Eq.~\eqref{stmass} and the PS mass function Eq.~\eqref{press}, for $\beta=0$ (solid), $\beta=0.05$ (dashed) and $\beta=0.08$ (dotted).}
\end{figure}

In the right panel of Fig.~\ref{numberf} we display the comoving number counts of dark matter halos for the Sheth-Tormen mass functions, for the coupling values of $\beta=0$, $\beta=0.05$ and $\beta=0.08$. The impact of the coupling follows the same trend as in the PS formalism, however, more pronnouced: for example, the suppression of the number counts at low redshifts can be prolonged up to higher $z$. As it is known \cite{Devi:2011gb} the ST mass function suppresses the number of objects at low redshifts and enhances that number at high redshifts when comparing to the PS formalism. This trend can be better seen in Fig.~\ref{stfig2} where we show the number difference between the PS and the ST formalisms. The discrepancy between the two mass functions peaks at  $z\sim 1$ where the difference can reach $\sim 2\times 10^5$ clusters (for $\Lambda$CDM).

Although the ST model gives a better fit to numerical simulations of the distribution of halos compared to the standard PS, N-body simulations have been able to find improved fitting mass functions \cite{Bhattacharya:2010wy} for a wide variety of cosmologies. However, for the scope of this work either the PS or the ST functions attend our purposes.

\section{Observations}
\label{observations}

The importance of linking the theory predictions with observational data leads us to this present section. In regard to our previous analysis, we follow to compute the predicted number of cluster-sized objects for two separate surveys and shed some light on its ability to distinguish between theoretical models. Prior studies have implemented a similar analysis for different theoretical models, such as disformally coupled \cite{Sapa:2018jja}, thawing \cite{Devi:2011gb} and freezing \cite{Devi:2014rva} models of dark energy.

\begin{figure*}[t]
\begin{center}
\includegraphics[width=1\textwidth]{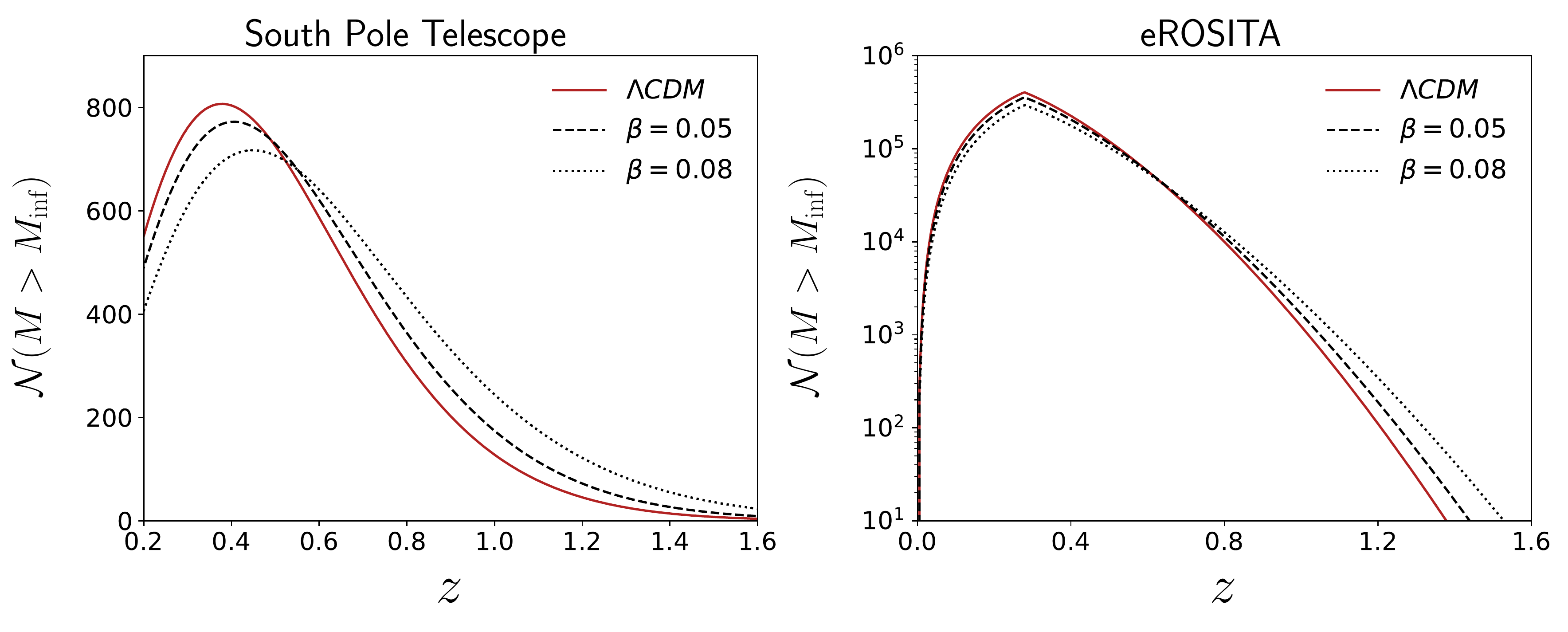}
\end{center}
\caption{\label{er1} Estimate on the number of galaxy clusters for the SPT SZ survey (left) and expected number of clusters along redshift for the  eROSITA (right), for $\beta=0$ (solid), $\beta=0.05$ (dashed) and $\beta=0.08$ (dotted).}
\end{figure*}

The 10 meter South Pole Telescope \cite{Carlstrom:2009um} is conducting a survey of galaxy clusters on the southern hemisphere sky. At present it is operating with its third-generation camera SPT-3G, but we focus our estimate on its first survey SPT-SZ \cite{Bleem:2014iim}. This mission covered an area of 2500 deg$^2$ (corresponding to a fraction of the sky of $f_{\text{sky}}\approx 0.06$) using the Sunyaev-Zel'dovich (SZ) Effect \cite{Sunyaev:1972eq}, from 2007 until 2011. The observational strategy can be found in \cite{Schaffer:2011mz} and the criteria for cluster selection, determination of redshift and other characteristics of the survey can be found in \cite{Bleem:2014iim}. Cosmological constraints from the SPT-SZ survey were conducted in \cite{deHaan:2016qvy}, considering a sample of clusters at $z>0.25$. Our aim here is to estimate the effect of our $\beta$ coupling parameter on the number of observed galaxy clusters $\mathcal{N}$.

Following the SPT-SZ survey criteria in \cite{deHaan:2016qvy}, we use the detection significance $\xi$ parameter as an estimate of the cluster mass. More specifically, the cluster mass is estimated using the unbiased significance $\zeta$ related to $\xi$ through $\zeta = \sqrt{\langle\xi\rangle^2 - 3}$.  The mass scaling relation is then parametrized through
\begin{equation}
\label{sptscaling}
\zeta = A \left( \frac{M}{3\times 10^{14}M_{\odot}h^{-1}} \right)^B \left( \frac{E(z)}{E(0.6)} \right)^C,
\end{equation}
where  $E(z) = H(z)/H(0)$. $A$, $B$ and $C$ are parameters ultimately fitted by the data. Here we use $(A,B,C)=(3.531,1.661,0.733)$  (SPT+Planck+WP+BAO \cite{deHaan:2016qvy}). Imposing the selection criteria used in the SPT-SZ survey of $\xi > 5$, we follow to solve Eq.\eqref{number} with the integration being performed from $M_{\rm inf}=$ max$\left[M_l,10^{14}M_{\odot}\right]$, where $M_l$ is the mass limit obtained by solving Eq.~\eqref{sptscaling} for $M$. Recall that the result has to be multiplied by the fraction of the sky covered by the survey, $f_\text{sky}  = 0.06$, to only capture the objects within that region.  In our numerical simulations we integrate up to a mass of $M_{\text{sup}}=10^{16}M_{\odot}$ as  no structures are expected to form with larger masses. Nonetheless, we verified that increasing this upper bound does not affect our results.

In the left panel of Fig.~\ref{er1} we report the estimated number of galaxy clusters for the SPT-SZ survey as a function of redshift, using the Press-Schechter mass function for $\Lambda$CDM, $\beta=0.05$ and $\beta=0.08$. The values for $\Lambda$CDM are of the same order as the ones found in \cite{deHaan:2016qvy,Bocquet:2018ukq}. We expect that the number of detected galaxy clusters peaks at $z\approx 0.4$ with a value up to $\approx 800$. This number is slightly suppressed if a coupling is present. An interaction between dark energy and dark matter leaves an evident signature in the spectrum of $\mathcal{N}$.

\begin{figure*}[t]
\begin{center}
\includegraphics[width=1\textwidth]{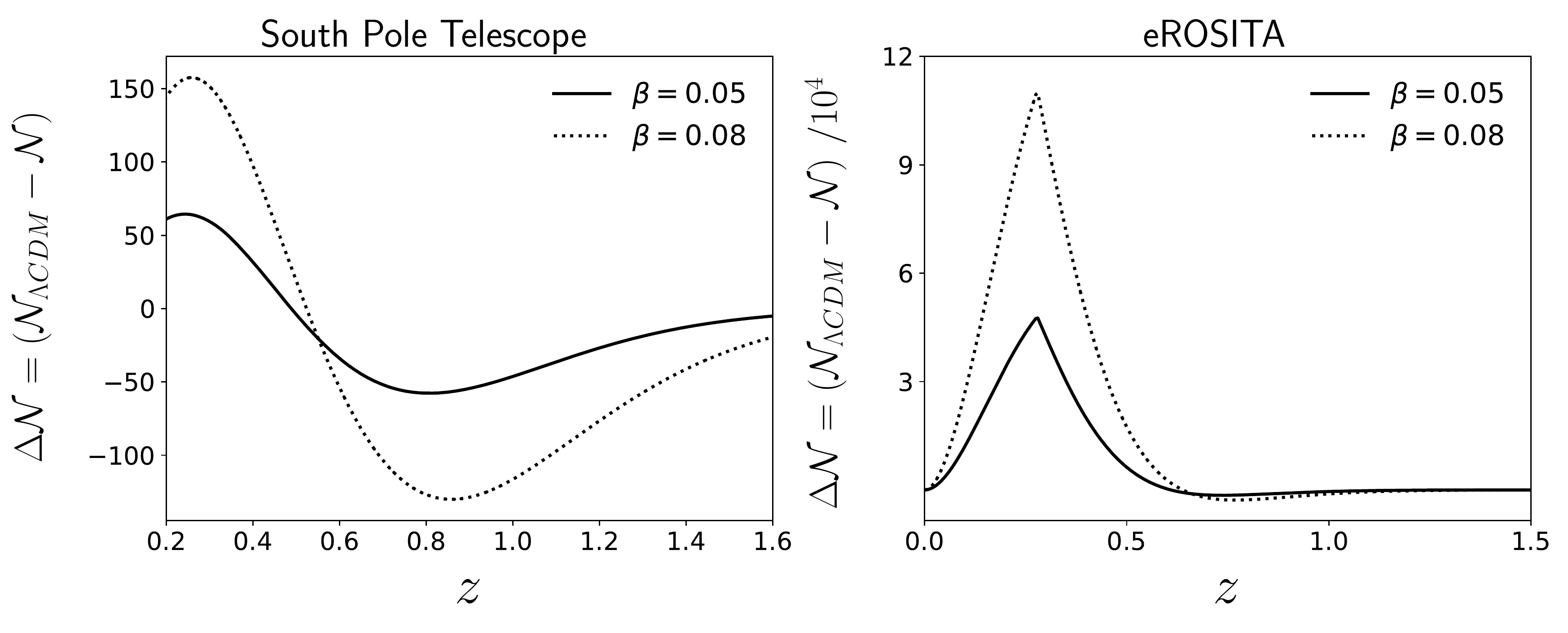}
\end{center}
\caption{\label{er2} Difference on the expected number of galaxy clusters from $\Lambda$CDM, for the SPT (left) and eROSITA (right) surveys, for $\beta=0.05$ (dashed) and $\beta=0.08$ (dotted).}
\end{figure*}

It is also crucial to analyze if the differences for the coupled models to the standard $\Lambda$CDM are within the range in which the survey will be able to discriminate. This difference is shown in the left panel of Fig.~\ref{er2}. Clearly there is a  discrepancy on the number counts predicted by the different models. This difference peaks around $z\approx 0.3$, where $\Delta\mathcal{N}\approx 150$ for $\beta = 0.08$ and $\Delta\mathcal{N}\approx 60$ for $\beta = 0.05$. These values are  above the estimated SPT uncertainty $\Delta\mathcal{N}\approx 50$ \cite{Vanderlinde_2010,Story:2011cr}. Hence, this suggests that in principle it would be possible to distinguish between $\Lambda$CDM and coupled quintessence models with the  SPT-SZ survey. However this only holds if one assumes that all the remaining cosmological parameters are determined. In particular the values of $\mathcal{N}$ have a strong dependence on $\sigma_8$ as seen in Fig.\ref{sigmaPlot}.

The second survey that we shall address here is the X-ray telescope eROSITA. Compared with the SPT, it covers a much wider fraction of the sky, $f_{\text{sky}}= 0.485$. The limiting energy flux in the band $[0.5,2.0]$ KeV is $f_{lim} = 3.3\times 10^{-14}$ erg s$^{-1}$ cm$^{-2}$. To convert the limiting flux to a mass, in order to perform the integration of the expected number counts, we follow the procedure of \cite{Basilakos:2010fb,Fedeli:2009fj}. Specifically, the relation between bolometric X-ray luminosity and mass can be written as
\begin{equation}
\label{eRosLM}
L(M,z) = 3.087\times 10^{44}\left[ \frac{M\,E(z)}{10^{15}h^{-1}M_{\odot}} \right]^{1.554}h^{-2}.
\end{equation} 
Analogous to the previous survey, we then find the limiting mass by solving Eq.~\eqref{eRosLM} in $M$ with the luminosity given by $L=4\pi d_L^2f_{\lim}c_b$. The parameter $c_b$ is a band correction necessary to convert from a bolometric luminosity into the eROSITA energy band \cite{Fedeli:2009fj,Fedeli:2008fh}. 
In this work we set $c_b \approx 1.5$.

The value of the limiting mass with redshift for several dynamical dark energy cosmologies in the context of both surveys can be found in \cite{Fedeli:2008fh}.

The results for the expected halo number counts is depicted in the right panel of Fig.~\ref{er1} and in the right panel of Fig.~\ref{er2} we show the difference on the halo counts relative to the standard $\Lambda$CDM. Clearly eROSITA is expected to measure a higher number of clusters than the SPT, peaking at $z\approx 0.28$ where the differences relative to $\Lambda$CDM, at this redshift, are $\Delta\mathcal{N}\approx 47\,000$ and $\Delta\mathcal{N}\approx 110\,000$ for $\beta = 0.05$ and $\beta = 0.08$, respectively. These numbers are much higher than the expected eROSITA sensitivity of $\Delta \mathcal{N}\approx 500$ \cite{Merloni:2012uf,Devi:2014rva} suggesting the possibility to discriminate between models with eROSITA. The values found for $\mathcal{N}$ with $\beta = 0$ ($\Lambda$CDM) are, as expected, consistent with prior studies \cite{Sapa:2018jja,Basilakos:2010fb}.

It is worth mentioning that the forecast method applied in this section, for the eROSITA survey, was conducted using the Press-Schechter mass function. Nonetheless, we have numerically verified that the ellipsoidal collapse model of Sheth-Tormen, described in Sec.~\ref{st} holds the same conclusions of the spherical model, with values of $\Delta\mathcal{N}$ being remarkably close to the ones found for PS in Fig.~\ref{er2}.

\vspace{0.7cm}

A last remark that should be addressed is the fact that we have chosen not to include baryons throughout this work. At this level it is reasonable to assume that the baryons only evolve as dust. Its main effect is to add a separate uncoupled species of matter, thus diluting the impact of the coupling.  By adding the baryon fluid the reference value for the parameters may change slightly, but overall our main conclusions will remain unaltered.

\section{Conclusions}
\label{conclusions}

In this work we  discussed a coupled quintessence model tailored to mirror a $\Lambda$CDM evolution at background level, having explored the particular imprints left by the interaction on the spherical collapse parameters and on cluster abundances. We have shown that the transfer of energy from the dark matter component into the quintessence field leads to a slower growth of the matter density contrast. In other words, the perturbation evolves longer before the collapse takes place and, consequently, we obtain higher values for the critical density contrast $\delta_c$ as the coupling $\beta$ increases.

We used two different mass functions to compute the halo number counts. In both cases we obtain an enhancement on the number of objects at high redshifts and a suppression at small redshifts compared to the standard $\Lambda$CDM scenario. We also estimated the expected number of clusters that can be observed in two  surveys -- the eROSITA satellite mission and the South Pole Telescope SZ survey. In principle both these missions, but in particular the eROSITA one, should be able to distinguish between $\Lambda$CDM and a model with nonzero coupling. Of course the result is also highly dependent on other cosmological parameters, and in particular $\sigma_8$, so that the degeneracy can only be lifted with a precise determination of these parameters from independent observations. Bear also in mind that allowed values of the coupling  can also depend on the value of $\sigma_8$ \cite{Barros:2018efl}.
One distinct feature of this model, in contrast with a standard coupled quintessence, is the fact that the volume element present on the cluster spectrum Eq.~\eqref{number} and the limiting mass of each treated survey do not vary with the coupling and have  the same values as $\Lambda$CDM. This is due to the fact that both these quantities depend only on the background function $H(z)$, which here is settled to evolve as a $\Lambda$CDM cosmology.

With this work we see that the analysis of the nonlinear collapse in this coupled dark energy model is a promising way of testing it against the standard $\Lambda$CDM model. This result will have to be confirmed with a more rigorous analysis, for instance resorting to  N-body simulations as was carried out in \cite{Baldi:2008ay} for the usual coupled quintessence models.

\acknowledgments

The authors acknowledge the financial support by Funda\c{c}\~ao para a Ci\^encia e a Tecnologia (FCT) through the research grants: UID/FIS/04434/2019, PTDC/FIS-OUT/29048/2017 (DarkRipple),  COMPETE2020: POCI-01-0145-FEDER-028987 \& FCT: PTDC/FIS-AST/28987/2017 (CosmoESPRESSO) and IF/00852/2015 (Dark Couplings). B.J.B is supported by the grant PD/BD/128018/2016 (PhD::SPACE program) from Funda\c{c}\~ao para a Ci\^encia e Tecnologia.

\bibliography{bib1}

\begin{thebibliography}{87}%
\makeatletter
\providecommand \@ifxundefined [1]{%
 \@ifx{#1\undefined}
}%
\providecommand \@ifnum [1]{%
 \ifnum #1\expandafter \@firstoftwo
 \else \expandafter \@secondoftwo
 \fi
}%
\providecommand \@ifx [1]{%
 \ifx #1\expandafter \@firstoftwo
 \else \expandafter \@secondoftwo
 \fi
}%
\providecommand \natexlab [1]{#1}%
\providecommand \enquote  [1]{``#1''}%
\providecommand \bibnamefont  [1]{#1}%
\providecommand \bibfnamefont [1]{#1}%
\providecommand \citenamefont [1]{#1}%
\providecommand \href@noop [0]{\@secondoftwo}%
\providecommand \href [0]{\begingroup \@sanitize@url \@href}%
\providecommand \@href[1]{\@@startlink{#1}\@@href}%
\providecommand \@@href[1]{\endgroup#1\@@endlink}%
\providecommand \@sanitize@url [0]{\catcode `\\12\catcode `\$12\catcode
  `\&12\catcode `\#12\catcode `\^12\catcode `\_12\catcode `\%12\relax}%
\providecommand \@@startlink[1]{}%
\providecommand \@@endlink[0]{}%
\providecommand \url  [0]{\begingroup\@sanitize@url \@url }%
\providecommand \@url [1]{\endgroup\@href {#1}{\urlprefix }}%
\providecommand \urlprefix  [0]{URL }%
\providecommand \Eprint [0]{\href }%
\providecommand \doibase [0]{http://dx.doi.org/}%
\providecommand \selectlanguage [0]{\@gobble}%
\providecommand \bibinfo  [0]{\@secondoftwo}%
\providecommand \bibfield  [0]{\@secondoftwo}%
\providecommand \translation [1]{[#1]}%
\providecommand \BibitemOpen [0]{}%
\providecommand \bibitemStop [0]{}%
\providecommand \bibitemNoStop [0]{.\EOS\space}%
\providecommand \EOS [0]{\spacefactor3000\relax}%
\providecommand \BibitemShut  [1]{\csname bibitem#1\endcsname}%
\let\auto@bib@innerbib\@empty
\bibitem [{\citenamefont {Jennings}(2012)}]{jennings2012}%
  \BibitemOpen
  \bibfield  {author} {\bibinfo {author} {\bibfnamefont {E.}~\bibnamefont
  {Jennings}},\ }\href {https://www.springer.com/gp/book/9783642293382} {\emph
  {\bibinfo {title} {Simulations of Dark Energy Cosmologies}}}\ (\bibinfo
  {publisher} {Springer-Verlag Berlin Heidelberg},\ \bibinfo {year}
  {2012})\BibitemShut {NoStop}%
\bibitem [{\citenamefont {Maccio}\ \emph {et~al.}(2004)\citenamefont {Maccio},
  \citenamefont {Quercellini}, \citenamefont {Mainini}, \citenamefont
  {Amendola},\ and\ \citenamefont {Bonometto}}]{Maccio:2003yk}%
  \BibitemOpen
  \bibfield  {author} {\bibinfo {author} {\bibfnamefont {A.~V.}\ \bibnamefont
  {Maccio}}, \bibinfo {author} {\bibfnamefont {C.}~\bibnamefont {Quercellini}},
  \bibinfo {author} {\bibfnamefont {R.}~\bibnamefont {Mainini}}, \bibinfo
  {author} {\bibfnamefont {L.}~\bibnamefont {Amendola}}, \ and\ \bibinfo
  {author} {\bibfnamefont {S.~A.}\ \bibnamefont {Bonometto}},\ }\href {\doibase
  10.1103/PhysRevD.69.123516} {\bibfield  {journal} {\bibinfo  {journal} {Phys.
  Rev.}\ }\textbf {\bibinfo {volume} {D69}},\ \bibinfo {pages} {123516}
  (\bibinfo {year} {2004})},\ \Eprint {http://arxiv.org/abs/astro-ph/0309671}
  {arXiv:astro-ph/0309671 [astro-ph]} \BibitemShut {NoStop}%
\bibitem [{\citenamefont {Baldi}(2011)}]{Baldi:2010vv}%
  \BibitemOpen
  \bibfield  {author} {\bibinfo {author} {\bibfnamefont {M.}~\bibnamefont
  {Baldi}},\ }\href {\doibase 10.1111/j.1365-2966.2010.17758.x} {\bibfield
  {journal} {\bibinfo  {journal} {Mon. Not. Roy. Astron. Soc.}\ }\textbf
  {\bibinfo {volume} {411}},\ \bibinfo {pages} {1077} (\bibinfo {year}
  {2011})},\ \Eprint {http://arxiv.org/abs/1005.2188} {arXiv:1005.2188
  [astro-ph.CO]} \BibitemShut {NoStop}%
\bibitem [{\citenamefont {De~Boni}\ \emph {et~al.}(2011)\citenamefont
  {De~Boni}, \citenamefont {Dolag}, \citenamefont {Ettori}, \citenamefont
  {Moscardini}, \citenamefont {Pettorino},\ and\ \citenamefont
  {Baccigalupi}}]{10.1111/j.1365-2966.2011.18894.x}%
  \BibitemOpen
  \bibfield  {author} {\bibinfo {author} {\bibfnamefont {C.}~\bibnamefont
  {De~Boni}}, \bibinfo {author} {\bibfnamefont {K.}~\bibnamefont {Dolag}},
  \bibinfo {author} {\bibfnamefont {S.}~\bibnamefont {Ettori}}, \bibinfo
  {author} {\bibfnamefont {L.}~\bibnamefont {Moscardini}}, \bibinfo {author}
  {\bibfnamefont {V.}~\bibnamefont {Pettorino}}, \ and\ \bibinfo {author}
  {\bibfnamefont {C.}~\bibnamefont {Baccigalupi}},\ }\href {\doibase
  10.1111/j.1365-2966.2011.18894.x} {\bibfield  {journal} {\bibinfo  {journal}
  {Monthly Notices of the Royal Astronomical Society}\ }\textbf {\bibinfo
  {volume} {415}},\ \bibinfo {pages} {2758} (\bibinfo {year} {2011})},\ \Eprint
  {http://arxiv.org/abs/http://oup.prod.sis.lan/mnras/article-pdf/415/3/2758/5980673/mnras0415-2758.pdf}
  {http://oup.prod.sis.lan/mnras/article-pdf/415/3/2758/5980673/mnras0415-2758.pdf}
  \BibitemShut {NoStop}%
\bibitem [{\citenamefont {Gunn}\ and\ \citenamefont
  {Gott}(1972)}]{Gunn:1972sv}%
  \BibitemOpen
  \bibfield  {author} {\bibinfo {author} {\bibfnamefont {J.~E.}\ \bibnamefont
  {Gunn}}\ and\ \bibinfo {author} {\bibfnamefont {J.~R.}\ \bibnamefont {Gott},
  \bibfnamefont {III}},\ }\href {\doibase 10.1086/151605} {\bibfield  {journal}
  {\bibinfo  {journal} {Astrophys. J.}\ }\textbf {\bibinfo {volume} {176}},\
  \bibinfo {pages} {1} (\bibinfo {year} {1972})}\BibitemShut {NoStop}%
\bibitem [{\citenamefont {Pace}\ \emph
  {et~al.}(2010{\natexlab{a}})\citenamefont {Pace}, \citenamefont {Waizmann},\
  and\ \citenamefont {Bartelmann}}]{10.1111/j.1365-2966.2010.16841.x}%
  \BibitemOpen
  \bibfield  {author} {\bibinfo {author} {\bibfnamefont {F.}~\bibnamefont
  {Pace}}, \bibinfo {author} {\bibfnamefont {J.-C.}\ \bibnamefont {Waizmann}},
  \ and\ \bibinfo {author} {\bibfnamefont {M.}~\bibnamefont {Bartelmann}},\
  }\href {\doibase 10.1111/j.1365-2966.2010.16841.x} {\bibfield  {journal}
  {\bibinfo  {journal} {Monthly Notices of the Royal Astronomical Society}\
  }\textbf {\bibinfo {volume} {406}},\ \bibinfo {pages} {1865} (\bibinfo {year}
  {2010}{\natexlab{a}})},\ \Eprint
  {http://arxiv.org/abs/http://oup.prod.sis.lan/mnras/article-pdf/406/3/1865/3007639/mnras0406-1865.pdf}
  {http://oup.prod.sis.lan/mnras/article-pdf/406/3/1865/3007639/mnras0406-1865.pdf}
  \BibitemShut {NoStop}%
\bibitem [{\citenamefont {Padmanabhan}(1999)}]{Padmanabhan1999}%
  \BibitemOpen
  \bibfield  {author} {\bibinfo {author} {\bibfnamefont {T.}~\bibnamefont
  {Padmanabhan}},\ }\href@noop {} {\emph {\bibinfo {title} {Structure Formation
  in the Universe}}}\ (\bibinfo  {publisher} {Cambridge University Press},\
  \bibinfo {year} {1999})\BibitemShut {NoStop}%
\bibitem [{\citenamefont {Liddle}\ and\ \citenamefont
  {Lyth}(1993)}]{Liddle:1993fq}%
  \BibitemOpen
  \bibfield  {author} {\bibinfo {author} {\bibfnamefont {A.~R.}\ \bibnamefont
  {Liddle}}\ and\ \bibinfo {author} {\bibfnamefont {D.~H.}\ \bibnamefont
  {Lyth}},\ }\href {\doibase 10.1016/0370-1573(93)90114-S} {\bibfield
  {journal} {\bibinfo  {journal} {Phys. Rept.}\ }\textbf {\bibinfo {volume}
  {231}},\ \bibinfo {pages} {1} (\bibinfo {year} {1993})},\ \Eprint
  {http://arxiv.org/abs/astro-ph/9303019} {arXiv:astro-ph/9303019 [astro-ph]}
  \BibitemShut {NoStop}%
\bibitem [{\citenamefont {Wetterich}(1995)}]{Wetterich:1994bg}%
  \BibitemOpen
  \bibfield  {author} {\bibinfo {author} {\bibfnamefont {C.}~\bibnamefont
  {Wetterich}},\ }\href@noop {} {\bibfield  {journal} {\bibinfo  {journal}
  {Astron. Astrophys.}\ }\textbf {\bibinfo {volume} {301}},\ \bibinfo {pages}
  {321} (\bibinfo {year} {1995})},\ \Eprint
  {http://arxiv.org/abs/hep-th/9408025} {arXiv:hep-th/9408025 [hep-th]}
  \BibitemShut {NoStop}%
\bibitem [{\citenamefont {Amendola}(2000{\natexlab{a}})}]{Amendola:1999er}%
  \BibitemOpen
  \bibfield  {author} {\bibinfo {author} {\bibfnamefont {L.}~\bibnamefont
  {Amendola}},\ }\href {\doibase 10.1103/PhysRevD.62.043511} {\bibfield
  {journal} {\bibinfo  {journal} {Phys. Rev.}\ }\textbf {\bibinfo {volume}
  {D62}},\ \bibinfo {pages} {043511} (\bibinfo {year} {2000}{\natexlab{a}})},\
  \Eprint {http://arxiv.org/abs/astro-ph/9908023} {arXiv:astro-ph/9908023
  [astro-ph]} \BibitemShut {NoStop}%
\bibitem [{\citenamefont {Barreiro}\ \emph {et~al.}(2000)\citenamefont
  {Barreiro}, \citenamefont {Copeland},\ and\ \citenamefont
  {Nunes}}]{Barreiro:1999zs}%
  \BibitemOpen
  \bibfield  {author} {\bibinfo {author} {\bibfnamefont {T.}~\bibnamefont
  {Barreiro}}, \bibinfo {author} {\bibfnamefont {E.~J.}\ \bibnamefont
  {Copeland}}, \ and\ \bibinfo {author} {\bibfnamefont {N.~J.}\ \bibnamefont
  {Nunes}},\ }\href {\doibase 10.1103/PhysRevD.61.127301} {\bibfield  {journal}
  {\bibinfo  {journal} {Phys. Rev.}\ }\textbf {\bibinfo {volume} {D61}},\
  \bibinfo {pages} {127301} (\bibinfo {year} {2000})},\ \Eprint
  {http://arxiv.org/abs/astro-ph/9910214} {arXiv:astro-ph/9910214 [astro-ph]}
  \BibitemShut {NoStop}%
\bibitem [{\citenamefont {Zimdahl}\ and\ \citenamefont
  {Pavon}(2001)}]{Zimdahl:2001ar}%
  \BibitemOpen
  \bibfield  {author} {\bibinfo {author} {\bibfnamefont {W.}~\bibnamefont
  {Zimdahl}}\ and\ \bibinfo {author} {\bibfnamefont {D.}~\bibnamefont
  {Pavon}},\ }\href {\doibase 10.1016/S0370-2693(01)01174-1} {\bibfield
  {journal} {\bibinfo  {journal} {Phys. Lett.}\ }\textbf {\bibinfo {volume}
  {B521}},\ \bibinfo {pages} {133} (\bibinfo {year} {2001})},\ \Eprint
  {http://arxiv.org/abs/astro-ph/0105479} {arXiv:astro-ph/0105479 [astro-ph]}
  \BibitemShut {NoStop}%
\bibitem [{\citenamefont {Farrar}\ and\ \citenamefont
  {Peebles}(2004)}]{Farrar:2003uw}%
  \BibitemOpen
  \bibfield  {author} {\bibinfo {author} {\bibfnamefont {G.~R.}\ \bibnamefont
  {Farrar}}\ and\ \bibinfo {author} {\bibfnamefont {P.~J.~E.}\ \bibnamefont
  {Peebles}},\ }\href {\doibase 10.1086/381728} {\bibfield  {journal} {\bibinfo
   {journal} {Astrophys. J.}\ }\textbf {\bibinfo {volume} {604}},\ \bibinfo
  {pages} {1} (\bibinfo {year} {2004})},\ \Eprint
  {http://arxiv.org/abs/astro-ph/0307316} {arXiv:astro-ph/0307316 [astro-ph]}
  \BibitemShut {NoStop}%
\bibitem [{\citenamefont {Amendola}\ \emph {et~al.}(2008)\citenamefont
  {Amendola}, \citenamefont {Baldi},\ and\ \citenamefont
  {Wetterich}}]{Amendola:2007yx}%
  \BibitemOpen
  \bibfield  {author} {\bibinfo {author} {\bibfnamefont {L.}~\bibnamefont
  {Amendola}}, \bibinfo {author} {\bibfnamefont {M.}~\bibnamefont {Baldi}}, \
  and\ \bibinfo {author} {\bibfnamefont {C.}~\bibnamefont {Wetterich}},\ }\href
  {\doibase 10.1103/PhysRevD.78.023015} {\bibfield  {journal} {\bibinfo
  {journal} {Phys. Rev.}\ }\textbf {\bibinfo {volume} {D78}},\ \bibinfo {pages}
  {023015} (\bibinfo {year} {2008})},\ \Eprint {http://arxiv.org/abs/0706.3064}
  {arXiv:0706.3064 [astro-ph]} \BibitemShut {NoStop}%
\bibitem [{\citenamefont {Tsujikawa}(2011)}]{Tsujikawa:2010sc}%
  \BibitemOpen
  \bibfield  {author} {\bibinfo {author} {\bibfnamefont {S.}~\bibnamefont
  {Tsujikawa}},\ }\href {\doibase 10.1007/978-90-481-8685-3_8} {\ \textbf
  {\bibinfo {volume} {370}},\ \bibinfo {pages} {331} (\bibinfo {year}
  {2011})},\ \Eprint {http://arxiv.org/abs/1004.1493} {arXiv:1004.1493
  [astro-ph.CO]} \BibitemShut {NoStop}%
\bibitem [{\citenamefont {Copeland}\ \emph {et~al.}(2006)\citenamefont
  {Copeland}, \citenamefont {Sami},\ and\ \citenamefont
  {Tsujikawa}}]{Copeland:2006wr}%
  \BibitemOpen
  \bibfield  {author} {\bibinfo {author} {\bibfnamefont {E.~J.}\ \bibnamefont
  {Copeland}}, \bibinfo {author} {\bibfnamefont {M.}~\bibnamefont {Sami}}, \
  and\ \bibinfo {author} {\bibfnamefont {S.}~\bibnamefont {Tsujikawa}},\ }\href
  {\doibase 10.1142/S021827180600942X} {\bibfield  {journal} {\bibinfo
  {journal} {Int. J. Mod. Phys.}\ }\textbf {\bibinfo {volume} {D15}},\ \bibinfo
  {pages} {1753} (\bibinfo {year} {2006})},\ \Eprint
  {http://arxiv.org/abs/hep-th/0603057} {arXiv:hep-th/0603057 [hep-th]}
  \BibitemShut {NoStop}%
\bibitem [{\citenamefont {Luca~Amendola}(2010)}]{detao}%
  \BibitemOpen
  \bibfield  {author} {\bibinfo {author} {\bibfnamefont {S.~T.}\ \bibnamefont
  {Luca~Amendola}},\ }\href@noop {} {\emph {\bibinfo {title} {{Dark Energy:
  Theory and Observations}}}}\ (\bibinfo  {publisher} {Cambridge University
  Press},\ \bibinfo {year} {2010})\BibitemShut {NoStop}%
\bibitem [{\citenamefont {Wintergerst}\ and\ \citenamefont
  {Pettorino}(2010)}]{Wintergerst:2010ui}%
  \BibitemOpen
  \bibfield  {author} {\bibinfo {author} {\bibfnamefont {N.}~\bibnamefont
  {Wintergerst}}\ and\ \bibinfo {author} {\bibfnamefont {V.}~\bibnamefont
  {Pettorino}},\ }\href {\doibase 10.1103/PhysRevD.82.103516} {\bibfield
  {journal} {\bibinfo  {journal} {Phys. Rev.}\ }\textbf {\bibinfo {volume}
  {D82}},\ \bibinfo {pages} {103516} (\bibinfo {year} {2010})},\ \Eprint
  {http://arxiv.org/abs/1005.1278} {arXiv:1005.1278 [astro-ph.CO]} \BibitemShut
  {NoStop}%
\bibitem [{\citenamefont {Wintergerst}\ \emph {et~al.}(2010)\citenamefont
  {Wintergerst}, \citenamefont {Pettorino}, \citenamefont {Mota},\ and\
  \citenamefont {Wetterich}}]{Wintergerst:2009fh}%
  \BibitemOpen
  \bibfield  {author} {\bibinfo {author} {\bibfnamefont {N.}~\bibnamefont
  {Wintergerst}}, \bibinfo {author} {\bibfnamefont {V.}~\bibnamefont
  {Pettorino}}, \bibinfo {author} {\bibfnamefont {D.~F.}\ \bibnamefont {Mota}},
  \ and\ \bibinfo {author} {\bibfnamefont {C.}~\bibnamefont {Wetterich}},\
  }\href {\doibase 10.1103/PhysRevD.81.063525} {\bibfield  {journal} {\bibinfo
  {journal} {Phys. Rev.}\ }\textbf {\bibinfo {volume} {D81}},\ \bibinfo {pages}
  {063525} (\bibinfo {year} {2010})},\ \Eprint {http://arxiv.org/abs/0910.4985}
  {arXiv:0910.4985 [astro-ph.CO]} \BibitemShut {NoStop}%
\bibitem [{\citenamefont {Pace}\ \emph {et~al.}(2014)\citenamefont {Pace},
  \citenamefont {Moscardini}, \citenamefont {Crittenden}, \citenamefont
  {Bartelmann},\ and\ \citenamefont {Pettorino}}]{Pace:2013pea}%
  \BibitemOpen
  \bibfield  {author} {\bibinfo {author} {\bibfnamefont {F.}~\bibnamefont
  {Pace}}, \bibinfo {author} {\bibfnamefont {L.}~\bibnamefont {Moscardini}},
  \bibinfo {author} {\bibfnamefont {R.}~\bibnamefont {Crittenden}}, \bibinfo
  {author} {\bibfnamefont {M.}~\bibnamefont {Bartelmann}}, \ and\ \bibinfo
  {author} {\bibfnamefont {V.}~\bibnamefont {Pettorino}},\ }\href {\doibase
  10.1093/mnras/stt1907} {\bibfield  {journal} {\bibinfo  {journal} {Mon. Not.
  Roy. Astron. Soc.}\ }\textbf {\bibinfo {volume} {437}},\ \bibinfo {pages}
  {547} (\bibinfo {year} {2014})},\ \Eprint {http://arxiv.org/abs/1307.7026}
  {arXiv:1307.7026 [astro-ph.CO]} \BibitemShut {NoStop}%
\bibitem [{\citenamefont {Le~Delliou}\ and\ \citenamefont
  {Barreiro}(2013)}]{Delliou:2012ik}%
  \BibitemOpen
  \bibfield  {author} {\bibinfo {author} {\bibfnamefont {M.}~\bibnamefont
  {Le~Delliou}}\ and\ \bibinfo {author} {\bibfnamefont {T.}~\bibnamefont
  {Barreiro}},\ }\href {\doibase 10.1088/1475-7516/2013/02/037} {\bibfield
  {journal} {\bibinfo  {journal} {JCAP}\ }\textbf {\bibinfo {volume} {1302}},\
  \bibinfo {pages} {037} (\bibinfo {year} {2013})},\ \Eprint
  {http://arxiv.org/abs/1208.6373} {arXiv:1208.6373 [astro-ph.CO]} \BibitemShut
  {NoStop}%
\bibitem [{\citenamefont {Nazari-Pooya}\ \emph {et~al.}(2016)\citenamefont
  {Nazari-Pooya}, \citenamefont {Malekjani}, \citenamefont {Pace},\ and\
  \citenamefont {Jassur}}]{Nazari-Pooya:2016bra}%
  \BibitemOpen
  \bibfield  {author} {\bibinfo {author} {\bibfnamefont {N.}~\bibnamefont
  {Nazari-Pooya}}, \bibinfo {author} {\bibfnamefont {M.}~\bibnamefont
  {Malekjani}}, \bibinfo {author} {\bibfnamefont {F.}~\bibnamefont {Pace}}, \
  and\ \bibinfo {author} {\bibfnamefont {D.~M.-Z.}\ \bibnamefont {Jassur}},\
  }\href {\doibase 10.1093/mnras/stw582} {\bibfield  {journal} {\bibinfo
  {journal} {Mon. Not. Roy. Astron. Soc.}\ }\textbf {\bibinfo {volume} {458}},\
  \bibinfo {pages} {3795} (\bibinfo {year} {2016})},\ \Eprint
  {http://arxiv.org/abs/1601.04593} {arXiv:1601.04593 [gr-qc]} \BibitemShut
  {NoStop}%
\bibitem [{\citenamefont {Mota}(2008)}]{Mota:2008ne}%
  \BibitemOpen
  \bibfield  {author} {\bibinfo {author} {\bibfnamefont {D.~F.}\ \bibnamefont
  {Mota}},\ }\href {\doibase 10.1088/1475-7516/2008/09/006} {\bibfield
  {journal} {\bibinfo  {journal} {JCAP}\ }\textbf {\bibinfo {volume} {0809}},\
  \bibinfo {pages} {006} (\bibinfo {year} {2008})},\ \Eprint
  {http://arxiv.org/abs/0812.4493} {arXiv:0812.4493 [astro-ph]} \BibitemShut
  {NoStop}%
\bibitem [{\citenamefont {Mota}\ and\ \citenamefont {van~de
  Bruck}(2004)}]{Mota:2004pa}%
  \BibitemOpen
  \bibfield  {author} {\bibinfo {author} {\bibfnamefont {D.~F.}\ \bibnamefont
  {Mota}}\ and\ \bibinfo {author} {\bibfnamefont {C.}~\bibnamefont {van~de
  Bruck}},\ }\href {\doibase 10.1051/0004-6361:20041090} {\bibfield  {journal}
  {\bibinfo  {journal} {Astron. Astrophys.}\ }\textbf {\bibinfo {volume}
  {421}},\ \bibinfo {pages} {71} (\bibinfo {year} {2004})},\ \Eprint
  {http://arxiv.org/abs/astro-ph/0401504} {arXiv:astro-ph/0401504 [astro-ph]}
  \BibitemShut {NoStop}%
\bibitem [{\citenamefont {He}\ \emph {et~al.}(2010)\citenamefont {He},
  \citenamefont {Wang}, \citenamefont {Abdalla},\ and\ \citenamefont
  {Pavon}}]{He:2010ta}%
  \BibitemOpen
  \bibfield  {author} {\bibinfo {author} {\bibfnamefont {J.-H.}\ \bibnamefont
  {He}}, \bibinfo {author} {\bibfnamefont {B.}~\bibnamefont {Wang}}, \bibinfo
  {author} {\bibfnamefont {E.}~\bibnamefont {Abdalla}}, \ and\ \bibinfo
  {author} {\bibfnamefont {D.}~\bibnamefont {Pavon}},\ }\href {\doibase
  10.1088/1475-7516/2010/12/022} {\bibfield  {journal} {\bibinfo  {journal}
  {JCAP}\ }\textbf {\bibinfo {volume} {1012}},\ \bibinfo {pages} {022}
  (\bibinfo {year} {2010})},\ \Eprint {http://arxiv.org/abs/1001.0079}
  {arXiv:1001.0079 [gr-qc]} \BibitemShut {NoStop}%
\bibitem [{\citenamefont {Nunes}\ \emph {et~al.}(2006)\citenamefont {Nunes},
  \citenamefont {da~Silva},\ and\ \citenamefont {Aghanim}}]{Nunes:2005fn}%
  \BibitemOpen
  \bibfield  {author} {\bibinfo {author} {\bibfnamefont {N.~J.}\ \bibnamefont
  {Nunes}}, \bibinfo {author} {\bibfnamefont {A.~C.}\ \bibnamefont {da~Silva}},
  \ and\ \bibinfo {author} {\bibfnamefont {N.}~\bibnamefont {Aghanim}},\ }\href
  {\doibase 10.1051/0004-6361:20053706} {\bibfield  {journal} {\bibinfo
  {journal} {Astron. Astrophys.}\ }\textbf {\bibinfo {volume} {450}},\ \bibinfo
  {pages} {899} (\bibinfo {year} {2006})},\ \Eprint
  {http://arxiv.org/abs/astro-ph/0506043} {arXiv:astro-ph/0506043 [astro-ph]}
  \BibitemShut {NoStop}%
\bibitem [{\citenamefont {Peebles}\ and\ \citenamefont
  {Ratra}(2003)}]{Peebles:2002gy}%
  \BibitemOpen
  \bibfield  {author} {\bibinfo {author} {\bibfnamefont {P.~J.~E.}\
  \bibnamefont {Peebles}}\ and\ \bibinfo {author} {\bibfnamefont
  {B.}~\bibnamefont {Ratra}},\ }\href {\doibase 10.1103/RevModPhys.75.559}
  {\bibfield  {journal} {\bibinfo  {journal} {Rev. Mod. Phys.}\ }\textbf
  {\bibinfo {volume} {75}},\ \bibinfo {pages} {559} (\bibinfo {year} {2003})},\
  \Eprint {http://arxiv.org/abs/astro-ph/0207347} {arXiv:astro-ph/0207347
  [astro-ph]} \BibitemShut {NoStop}%
\bibitem [{\citenamefont {Weinberg}(2000)}]{Weinberg:2000yb}%
  \BibitemOpen
  \bibfield  {author} {\bibinfo {author} {\bibfnamefont {S.}~\bibnamefont
  {Weinberg}},\ }in\ \href
  {http://www.slac.stanford.edu/spires/find/books/www?cl=QB461:I57:2000} {\emph
  {\bibinfo {booktitle} {{Sources and detection of dark matter and dark energy
  in the universe. Proceedings, 4th International Symposium, DM 2000, Marina
  del Rey, USA, February 23-25, 2000}}}}\ (\bibinfo {year} {2000})\ pp.\
  \bibinfo {pages} {18--26},\ \Eprint {http://arxiv.org/abs/astro-ph/0005265}
  {arXiv:astro-ph/0005265 [astro-ph]} \BibitemShut {NoStop}%
\bibitem [{\citenamefont {Zlatev}\ \emph
  {et~al.}(1999{\natexlab{a}})\citenamefont {Zlatev}, \citenamefont {Wang},\
  and\ \citenamefont {Steinhardt}}]{Zlatev:1998tr}%
  \BibitemOpen
  \bibfield  {author} {\bibinfo {author} {\bibfnamefont {I.}~\bibnamefont
  {Zlatev}}, \bibinfo {author} {\bibfnamefont {L.-M.}\ \bibnamefont {Wang}}, \
  and\ \bibinfo {author} {\bibfnamefont {P.~J.}\ \bibnamefont {Steinhardt}},\
  }\href {\doibase 10.1103/PhysRevLett.82.896} {\bibfield  {journal} {\bibinfo
  {journal} {Phys. Rev. Lett.}\ }\textbf {\bibinfo {volume} {82}},\ \bibinfo
  {pages} {896} (\bibinfo {year} {1999}{\natexlab{a}})},\ \Eprint
  {http://arxiv.org/abs/astro-ph/9807002} {arXiv:astro-ph/9807002 [astro-ph]}
  \BibitemShut {NoStop}%
\bibitem [{\citenamefont {Chimento}\ \emph
  {et~al.}(2003{\natexlab{a}})\citenamefont {Chimento}, \citenamefont {Jakubi},
  \citenamefont {Pavon},\ and\ \citenamefont {Zimdahl}}]{Chimento2003}%
  \BibitemOpen
  \bibfield  {author} {\bibinfo {author} {\bibfnamefont {L.~P.}\ \bibnamefont
  {Chimento}}, \bibinfo {author} {\bibfnamefont {A.~S.}\ \bibnamefont
  {Jakubi}}, \bibinfo {author} {\bibfnamefont {D.}~\bibnamefont {Pavon}}, \
  and\ \bibinfo {author} {\bibfnamefont {W.}~\bibnamefont {Zimdahl}},\ }\href
  {\doibase 10.1103/PhysRevD.67.083513} {\bibfield  {journal} {\bibinfo
  {journal} {Phys. Rev.}\ }\textbf {\bibinfo {volume} {D67}},\ \bibinfo {pages}
  {083513} (\bibinfo {year} {2003}{\natexlab{a}})},\ \Eprint
  {http://arxiv.org/abs/astro-ph/0303145} {arXiv:astro-ph/0303145 [astro-ph]}
  \BibitemShut {NoStop}%
\bibitem [{\citenamefont {Zlatev}\ \emph
  {et~al.}(1999{\natexlab{b}})\citenamefont {Zlatev}, \citenamefont {Wang},\
  and\ \citenamefont {Steinhardt}}]{Zlatev1999}%
  \BibitemOpen
  \bibfield  {author} {\bibinfo {author} {\bibfnamefont {I.}~\bibnamefont
  {Zlatev}}, \bibinfo {author} {\bibfnamefont {L.-M.}\ \bibnamefont {Wang}}, \
  and\ \bibinfo {author} {\bibfnamefont {P.~J.}\ \bibnamefont {Steinhardt}},\
  }\href {\doibase 10.1103/PhysRevLett.82.896} {\bibfield  {journal} {\bibinfo
  {journal} {Phys. Rev. Lett.}\ }\textbf {\bibinfo {volume} {82}},\ \bibinfo
  {pages} {896} (\bibinfo {year} {1999}{\natexlab{b}})},\ \Eprint
  {http://arxiv.org/abs/astro-ph/9807002} {arXiv:astro-ph/9807002 [astro-ph]}
  \BibitemShut {NoStop}%
\bibitem [{\citenamefont {Cai}\ and\ \citenamefont {Wang}(2005)}]{Cai:2004dk}%
  \BibitemOpen
  \bibfield  {author} {\bibinfo {author} {\bibfnamefont {R.-G.}\ \bibnamefont
  {Cai}}\ and\ \bibinfo {author} {\bibfnamefont {A.}~\bibnamefont {Wang}},\
  }\href {\doibase 10.1088/1475-7516/2005/03/002} {\bibfield  {journal}
  {\bibinfo  {journal} {JCAP}\ }\textbf {\bibinfo {volume} {0503}},\ \bibinfo
  {pages} {002} (\bibinfo {year} {2005})},\ \Eprint
  {http://arxiv.org/abs/hep-th/0411025} {arXiv:hep-th/0411025 [hep-th]}
  \BibitemShut {NoStop}%
\bibitem [{\citenamefont {Chimento}\ \emph
  {et~al.}(2003{\natexlab{b}})\citenamefont {Chimento}, \citenamefont {Jakubi},
  \citenamefont {Pavon},\ and\ \citenamefont {Zimdahl}}]{Chimento:2003iea}%
  \BibitemOpen
  \bibfield  {author} {\bibinfo {author} {\bibfnamefont {L.~P.}\ \bibnamefont
  {Chimento}}, \bibinfo {author} {\bibfnamefont {A.~S.}\ \bibnamefont
  {Jakubi}}, \bibinfo {author} {\bibfnamefont {D.}~\bibnamefont {Pavon}}, \
  and\ \bibinfo {author} {\bibfnamefont {W.}~\bibnamefont {Zimdahl}},\ }\href
  {\doibase 10.1103/PhysRevD.67.083513} {\bibfield  {journal} {\bibinfo
  {journal} {Phys. Rev.}\ }\textbf {\bibinfo {volume} {D67}},\ \bibinfo {pages}
  {083513} (\bibinfo {year} {2003}{\natexlab{b}})},\ \Eprint
  {http://arxiv.org/abs/astro-ph/0303145} {arXiv:astro-ph/0303145 [astro-ph]}
  \BibitemShut {NoStop}%
\bibitem [{\citenamefont {Macaulay}\ \emph {et~al.}(2013)\citenamefont
  {Macaulay}, \citenamefont {Wehus},\ and\ \citenamefont
  {Eriksen}}]{Macaulay2013}%
  \BibitemOpen
  \bibfield  {author} {\bibinfo {author} {\bibfnamefont {E.}~\bibnamefont
  {Macaulay}}, \bibinfo {author} {\bibfnamefont {I.~K.}\ \bibnamefont {Wehus}},
  \ and\ \bibinfo {author} {\bibfnamefont {H.~K.}\ \bibnamefont {Eriksen}},\
  }\href {\doibase 10.1103/PhysRevLett.111.161301} {\bibfield  {journal}
  {\bibinfo  {journal} {Phys. Rev. Lett.}\ }\textbf {\bibinfo {volume} {111}},\
  \bibinfo {pages} {161301} (\bibinfo {year} {2013})},\ \Eprint
  {http://arxiv.org/abs/1303.6583} {arXiv:1303.6583 [astro-ph.CO]} \BibitemShut
  {NoStop}%
\bibitem [{\citenamefont {Battye}\ \emph {et~al.}(2015)\citenamefont {Battye},
  \citenamefont {Charnock},\ and\ \citenamefont {Moss}}]{Battye:2014qga}%
  \BibitemOpen
  \bibfield  {author} {\bibinfo {author} {\bibfnamefont {R.~A.}\ \bibnamefont
  {Battye}}, \bibinfo {author} {\bibfnamefont {T.}~\bibnamefont {Charnock}}, \
  and\ \bibinfo {author} {\bibfnamefont {A.}~\bibnamefont {Moss}},\ }\href
  {\doibase 10.1103/PhysRevD.91.103508} {\bibfield  {journal} {\bibinfo
  {journal} {Phys. Rev.}\ }\textbf {\bibinfo {volume} {D91}},\ \bibinfo {pages}
  {103508} (\bibinfo {year} {2015})},\ \Eprint {http://arxiv.org/abs/1409.2769}
  {arXiv:1409.2769 [astro-ph.CO]} \BibitemShut {NoStop}%
\bibitem [{\citenamefont {Barros}\ \emph {et~al.}(2019)\citenamefont {Barros},
  \citenamefont {Amendola}, \citenamefont {Barreiro},\ and\ \citenamefont
  {Nunes}}]{Barros:2018efl}%
  \BibitemOpen
  \bibfield  {author} {\bibinfo {author} {\bibfnamefont {B.~J.}\ \bibnamefont
  {Barros}}, \bibinfo {author} {\bibfnamefont {L.}~\bibnamefont {Amendola}},
  \bibinfo {author} {\bibfnamefont {T.}~\bibnamefont {Barreiro}}, \ and\
  \bibinfo {author} {\bibfnamefont {N.~J.}\ \bibnamefont {Nunes}},\ }\href
  {\doibase 10.1088/1475-7516/2019/01/007} {\bibfield  {journal} {\bibinfo
  {journal} {JCAP}\ }\textbf {\bibinfo {volume} {1901}},\ \bibinfo {pages}
  {007} (\bibinfo {year} {2019})},\ \Eprint {http://arxiv.org/abs/1802.09216}
  {arXiv:1802.09216 [astro-ph.CO]} \BibitemShut {NoStop}%
\bibitem [{\citenamefont {Asghari}\ \emph {et~al.}(2019)\citenamefont
  {Asghari}, \citenamefont {Jiménez}, \citenamefont {Khosravi},\ and\
  \citenamefont {Mota}}]{Asghari:2019qld}%
  \BibitemOpen
  \bibfield  {author} {\bibinfo {author} {\bibfnamefont {M.}~\bibnamefont
  {Asghari}}, \bibinfo {author} {\bibfnamefont {J.~B.}\ \bibnamefont
  {Jiménez}}, \bibinfo {author} {\bibfnamefont {S.}~\bibnamefont {Khosravi}},
  \ and\ \bibinfo {author} {\bibfnamefont {D.~F.}\ \bibnamefont {Mota}},\
  }\href {\doibase 10.1088/1475-7516/2019/04/042} {\bibfield  {journal}
  {\bibinfo  {journal} {JCAP}\ }\textbf {\bibinfo {volume} {2019}},\ \bibinfo
  {pages} {042} (\bibinfo {year} {2019})},\ \Eprint
  {http://arxiv.org/abs/1902.05532} {arXiv:1902.05532 [astro-ph.CO]}
  \BibitemShut {NoStop}%
\bibitem [{\citenamefont {Simpson}(2010)}]{Simpson:2010vh}%
  \BibitemOpen
  \bibfield  {author} {\bibinfo {author} {\bibfnamefont {F.}~\bibnamefont
  {Simpson}},\ }\href {\doibase 10.1103/PhysRevD.82.083505} {\bibfield
  {journal} {\bibinfo  {journal} {Phys. Rev.}\ }\textbf {\bibinfo {volume}
  {D82}},\ \bibinfo {pages} {083505} (\bibinfo {year} {2010})},\ \Eprint
  {http://arxiv.org/abs/1007.1034} {arXiv:1007.1034 [astro-ph.CO]} \BibitemShut
  {NoStop}%
\bibitem [{\citenamefont {Baldi}\ and\ \citenamefont
  {Simpson}(2017)}]{Baldi:2016zom}%
  \BibitemOpen
  \bibfield  {author} {\bibinfo {author} {\bibfnamefont {M.}~\bibnamefont
  {Baldi}}\ and\ \bibinfo {author} {\bibfnamefont {F.}~\bibnamefont
  {Simpson}},\ }\href {\doibase 10.1093/mnras/stw2702} {\bibfield  {journal}
  {\bibinfo  {journal} {Mon. Not. Roy. Astron. Soc.}\ }\textbf {\bibinfo
  {volume} {465}},\ \bibinfo {pages} {653} (\bibinfo {year} {2017})},\ \Eprint
  {http://arxiv.org/abs/1605.05623} {arXiv:1605.05623 [astro-ph.CO]}
  \BibitemShut {NoStop}%
\bibitem [{\citenamefont {Kumar}\ and\ \citenamefont
  {Nunes}(2017)}]{Kumar:2017bpv}%
  \BibitemOpen
  \bibfield  {author} {\bibinfo {author} {\bibfnamefont {S.}~\bibnamefont
  {Kumar}}\ and\ \bibinfo {author} {\bibfnamefont {R.~C.}\ \bibnamefont
  {Nunes}},\ }\href {\doibase 10.1140/epjc/s10052-017-5334-3} {\bibfield
  {journal} {\bibinfo  {journal} {Eur. Phys. J.}\ }\textbf {\bibinfo {volume}
  {C77}},\ \bibinfo {pages} {734} (\bibinfo {year} {2017})},\ \Eprint
  {http://arxiv.org/abs/1709.02384} {arXiv:1709.02384 [astro-ph.CO]}
  \BibitemShut {NoStop}%
\bibitem [{\citenamefont {Aghanim}\ \emph {et~al.}(2018)\citenamefont {Aghanim}
  \emph {et~al.}}]{Aghanim:2018eyx}%
  \BibitemOpen
  \bibfield  {author} {\bibinfo {author} {\bibfnamefont {N.}~\bibnamefont
  {Aghanim}} \emph {et~al.} (\bibinfo {collaboration} {Planck}),\ }\href@noop
  {} {\  (\bibinfo {year} {2018})},\ \Eprint {http://arxiv.org/abs/1807.06209}
  {arXiv:1807.06209 [astro-ph.CO]} \BibitemShut {NoStop}%
\bibitem [{\citenamefont {Riess}\ \emph {et~al.}(2019)\citenamefont {Riess},
  \citenamefont {Casertano}, \citenamefont {Yuan}, \citenamefont {Macri},\ and\
  \citenamefont {Scolnic}}]{Riess:2019cxk}%
  \BibitemOpen
  \bibfield  {author} {\bibinfo {author} {\bibfnamefont {A.~G.}\ \bibnamefont
  {Riess}}, \bibinfo {author} {\bibfnamefont {S.}~\bibnamefont {Casertano}},
  \bibinfo {author} {\bibfnamefont {W.}~\bibnamefont {Yuan}}, \bibinfo {author}
  {\bibfnamefont {L.~M.}\ \bibnamefont {Macri}}, \ and\ \bibinfo {author}
  {\bibfnamefont {D.}~\bibnamefont {Scolnic}},\ }\href {\doibase
  10.3847/1538-4357/ab1422} {\bibfield  {journal} {\bibinfo  {journal}
  {Astrophys. J.}\ }\textbf {\bibinfo {volume} {876}},\ \bibinfo {pages} {85}
  (\bibinfo {year} {2019})},\ \Eprint {http://arxiv.org/abs/1903.07603}
  {arXiv:1903.07603 [astro-ph.CO]} \BibitemShut {NoStop}%
\bibitem [{\citenamefont {Merloni}\ \emph {et~al.}(2012)\citenamefont {Merloni}
  \emph {et~al.}}]{Merloni:2012uf}%
  \BibitemOpen
  \bibfield  {author} {\bibinfo {author} {\bibfnamefont {A.}~\bibnamefont
  {Merloni}} \emph {et~al.} (\bibinfo {collaboration} {eROSITA}),\ }\href@noop
  {} {\  (\bibinfo {year} {2012})},\ \Eprint {http://arxiv.org/abs/1209.3114}
  {arXiv:1209.3114 [astro-ph.HE]} \BibitemShut {NoStop}%
\bibitem [{\citenamefont {Pillepich}\ \emph {et~al.}(2012)\citenamefont
  {Pillepich}, \citenamefont {Porciani},\ and\ \citenamefont
  {Reiprich}}]{Pillepich:2011zz}%
  \BibitemOpen
  \bibfield  {author} {\bibinfo {author} {\bibfnamefont {A.}~\bibnamefont
  {Pillepich}}, \bibinfo {author} {\bibfnamefont {C.}~\bibnamefont {Porciani}},
  \ and\ \bibinfo {author} {\bibfnamefont {T.~H.}\ \bibnamefont {Reiprich}},\
  }\href {\doibase 10.1111/j.1365-2966.2012.20443.x} {\bibfield  {journal}
  {\bibinfo  {journal} {Mon. Not. Roy. Astron. Soc.}\ }\textbf {\bibinfo
  {volume} {422}},\ \bibinfo {pages} {44} (\bibinfo {year} {2012})},\ \Eprint
  {http://arxiv.org/abs/1111.6587} {arXiv:1111.6587 [astro-ph.CO]} \BibitemShut
  {NoStop}%
\bibitem [{\citenamefont {Ruhl}\ \emph {et~al.}(2004)\citenamefont {Ruhl} \emph
  {et~al.}}]{Ruhl:2004kv}%
  \BibitemOpen
  \bibfield  {author} {\bibinfo {author} {\bibfnamefont {J.~E.}\ \bibnamefont
  {Ruhl}} \emph {et~al.} (\bibinfo {collaboration} {SPT}),\ }\bibfield
  {booktitle} {\emph {\bibinfo {booktitle} {{SPIE Astronomical Telescopes and
  Instrumentation Symposium Glasgow, Scotland, United Kingdom, June 21-25,
  2004}}},\ }\href {\doibase 10.1117/12.552473} {\bibfield  {journal} {\bibinfo
   {journal} {Proc. SPIE Int. Soc. Opt. Eng.}\ }\textbf {\bibinfo {volume}
  {5498}},\ \bibinfo {pages} {11} (\bibinfo {year} {2004})},\ \Eprint
  {http://arxiv.org/abs/astro-ph/0411122} {arXiv:astro-ph/0411122 [astro-ph]}
  \BibitemShut {NoStop}%
\bibitem [{\citenamefont {de~Haan}\ \emph {et~al.}(2016)\citenamefont {de~Haan}
  \emph {et~al.}}]{deHaan:2016qvy}%
  \BibitemOpen
  \bibfield  {author} {\bibinfo {author} {\bibfnamefont {T.}~\bibnamefont
  {de~Haan}} \emph {et~al.} (\bibinfo {collaboration} {SPT}),\ }\href {\doibase
  10.3847/0004-637X/832/1/95} {\bibfield  {journal} {\bibinfo  {journal}
  {Astrophys. J.}\ }\textbf {\bibinfo {volume} {832}},\ \bibinfo {pages} {95}
  (\bibinfo {year} {2016})},\ \Eprint {http://arxiv.org/abs/1603.06522}
  {arXiv:1603.06522 [astro-ph.CO]} \BibitemShut {NoStop}%
\bibitem [{\citenamefont {Bleem}\ \emph {et~al.}(2015)\citenamefont {Bleem}
  \emph {et~al.}}]{Bleem:2014iim}%
  \BibitemOpen
  \bibfield  {author} {\bibinfo {author} {\bibfnamefont {L.~E.}\ \bibnamefont
  {Bleem}} \emph {et~al.} (\bibinfo {collaboration} {SPT}),\ }\href {\doibase
  10.1088/0067-0049/216/2/27} {\bibfield  {journal} {\bibinfo  {journal}
  {Astrophys. J. Suppl.}\ }\textbf {\bibinfo {volume} {216}},\ \bibinfo {pages}
  {27} (\bibinfo {year} {2015})},\ \Eprint {http://arxiv.org/abs/1409.0850}
  {arXiv:1409.0850 [astro-ph.CO]} \BibitemShut {NoStop}%
\bibitem [{\citenamefont {Barros}(2019)}]{Barros:2019rdv}%
  \BibitemOpen
  \bibfield  {author} {\bibinfo {author} {\bibfnamefont {B.~J.}\ \bibnamefont
  {Barros}},\ }\href {\doibase 10.1103/PhysRevD.99.064051} {\bibfield
  {journal} {\bibinfo  {journal} {Phys. Rev.}\ }\textbf {\bibinfo {volume}
  {D99}},\ \bibinfo {pages} {064051} (\bibinfo {year} {2019})},\ \Eprint
  {http://arxiv.org/abs/1901.03972} {arXiv:1901.03972 [gr-qc]} \BibitemShut
  {NoStop}%
\bibitem [{\citenamefont {Amendola}\ and\ \citenamefont
  {Tocchini-Valentini}(2002)}]{Amendola:2001rc}%
  \BibitemOpen
  \bibfield  {author} {\bibinfo {author} {\bibfnamefont {L.}~\bibnamefont
  {Amendola}}\ and\ \bibinfo {author} {\bibfnamefont {D.}~\bibnamefont
  {Tocchini-Valentini}},\ }\href {\doibase 10.1103/PhysRevD.66.043528}
  {\bibfield  {journal} {\bibinfo  {journal} {Phys. Rev.}\ }\textbf {\bibinfo
  {volume} {D66}},\ \bibinfo {pages} {043528} (\bibinfo {year} {2002})},\
  \Eprint {http://arxiv.org/abs/astro-ph/0111535} {arXiv:astro-ph/0111535
  [astro-ph]} \BibitemShut {NoStop}%
\bibitem [{\citenamefont {Amendola}(2004)}]{Amendola:2003wa}%
  \BibitemOpen
  \bibfield  {author} {\bibinfo {author} {\bibfnamefont {L.}~\bibnamefont
  {Amendola}},\ }\href {\doibase 10.1103/PhysRevD.69.103524} {\bibfield
  {journal} {\bibinfo  {journal} {Phys. Rev.}\ }\textbf {\bibinfo {volume}
  {D69}},\ \bibinfo {pages} {103524} (\bibinfo {year} {2004})},\ \Eprint
  {http://arxiv.org/abs/astro-ph/0311175} {arXiv:astro-ph/0311175 [astro-ph]}
  \BibitemShut {NoStop}%
\bibitem [{\citenamefont {Amendola}\ \emph {et~al.}(2014)\citenamefont
  {Amendola}, \citenamefont {Barreiro},\ and\ \citenamefont
  {Nunes}}]{Amendola:2014kwa}%
  \BibitemOpen
  \bibfield  {author} {\bibinfo {author} {\bibfnamefont {L.}~\bibnamefont
  {Amendola}}, \bibinfo {author} {\bibfnamefont {T.}~\bibnamefont {Barreiro}},
  \ and\ \bibinfo {author} {\bibfnamefont {N.~J.}\ \bibnamefont {Nunes}},\
  }\href {\doibase 10.1103/PhysRevD.90.083508} {\bibfield  {journal} {\bibinfo
  {journal} {Phys. Rev.}\ }\textbf {\bibinfo {volume} {D90}},\ \bibinfo {pages}
  {083508} (\bibinfo {year} {2014})},\ \Eprint {http://arxiv.org/abs/1407.2156}
  {arXiv:1407.2156 [astro-ph.CO]} \BibitemShut {NoStop}%
\bibitem [{\citenamefont {Teixeira}\ \emph {et~al.}(2019)\citenamefont
  {Teixeira}, \citenamefont {Nunes},\ and\ \citenamefont
  {Nunes}}]{Teixeira:2019tfi}%
  \BibitemOpen
  \bibfield  {author} {\bibinfo {author} {\bibfnamefont {E.~M.}\ \bibnamefont
  {Teixeira}}, \bibinfo {author} {\bibfnamefont {A.}~\bibnamefont {Nunes}}, \
  and\ \bibinfo {author} {\bibfnamefont {N.~J.}\ \bibnamefont {Nunes}},\
  }\href@noop {} {\  (\bibinfo {year} {2019})},\ \Eprint
  {http://arxiv.org/abs/1903.06028} {arXiv:1903.06028 [gr-qc]} \BibitemShut
  {NoStop}%
\bibitem [{\citenamefont {Amendola}(2000{\natexlab{b}})}]{Amendola:1999dr}%
  \BibitemOpen
  \bibfield  {author} {\bibinfo {author} {\bibfnamefont {L.}~\bibnamefont
  {Amendola}},\ }\href {\doibase 10.1046/j.1365-8711.2000.03165.x} {\bibfield
  {journal} {\bibinfo  {journal} {Mon. Not. Roy. Astron. Soc.}\ }\textbf
  {\bibinfo {volume} {312}},\ \bibinfo {pages} {521} (\bibinfo {year}
  {2000}{\natexlab{b}})},\ \Eprint {http://arxiv.org/abs/astro-ph/9906073}
  {arXiv:astro-ph/9906073 [astro-ph]} \BibitemShut {NoStop}%
\bibitem [{\citenamefont {Damour}\ \emph {et~al.}(1990)\citenamefont {Damour},
  \citenamefont {Gibbons},\ and\ \citenamefont
  {Gundlach}}]{PhysRevLett.64.123}%
  \BibitemOpen
  \bibfield  {author} {\bibinfo {author} {\bibfnamefont {T.}~\bibnamefont
  {Damour}}, \bibinfo {author} {\bibfnamefont {G.~W.}\ \bibnamefont {Gibbons}},
  \ and\ \bibinfo {author} {\bibfnamefont {C.}~\bibnamefont {Gundlach}},\
  }\href {\doibase 10.1103/PhysRevLett.64.123} {\bibfield  {journal} {\bibinfo
  {journal} {Phys. Rev. Lett.}\ }\textbf {\bibinfo {volume} {64}},\ \bibinfo
  {pages} {123} (\bibinfo {year} {1990})}\BibitemShut {NoStop}%
\bibitem [{\citenamefont {Koivisto}(2005)}]{Koivisto:2005nr}%
  \BibitemOpen
  \bibfield  {author} {\bibinfo {author} {\bibfnamefont {T.}~\bibnamefont
  {Koivisto}},\ }\href {\doibase 10.1103/PhysRevD.72.043516} {\bibfield
  {journal} {\bibinfo  {journal} {Phys. Rev.}\ }\textbf {\bibinfo {volume}
  {D72}},\ \bibinfo {pages} {043516} (\bibinfo {year} {2005})},\ \Eprint
  {http://arxiv.org/abs/astro-ph/0504571} {arXiv:astro-ph/0504571 [astro-ph]}
  \BibitemShut {NoStop}%
\bibitem [{\citenamefont {Bean}\ and\ \citenamefont
  {Magueijo}(2001)}]{Bean:2000zm}%
  \BibitemOpen
  \bibfield  {author} {\bibinfo {author} {\bibfnamefont {R.}~\bibnamefont
  {Bean}}\ and\ \bibinfo {author} {\bibfnamefont {J.}~\bibnamefont
  {Magueijo}},\ }\href {\doibase 10.1016/S0370-2693(01)00966-2} {\bibfield
  {journal} {\bibinfo  {journal} {Phys. Lett.}\ }\textbf {\bibinfo {volume}
  {B517}},\ \bibinfo {pages} {177} (\bibinfo {year} {2001})},\ \Eprint
  {http://arxiv.org/abs/astro-ph/0007199} {arXiv:astro-ph/0007199 [astro-ph]}
  \BibitemShut {NoStop}%
\bibitem [{\citenamefont {Liddle}\ \emph {et~al.}(1996)\citenamefont {Liddle},
  \citenamefont {Lyth}, \citenamefont {Schaefer}, \citenamefont {Shafi},\ and\
  \citenamefont {Viana}}]{Liddle:1995ay}%
  \BibitemOpen
  \bibfield  {author} {\bibinfo {author} {\bibfnamefont {A.~R.}\ \bibnamefont
  {Liddle}}, \bibinfo {author} {\bibfnamefont {D.~H.}\ \bibnamefont {Lyth}},
  \bibinfo {author} {\bibfnamefont {R.~K.}\ \bibnamefont {Schaefer}}, \bibinfo
  {author} {\bibfnamefont {Q.}~\bibnamefont {Shafi}}, \ and\ \bibinfo {author}
  {\bibfnamefont {P.~T.~P.}\ \bibnamefont {Viana}},\ }\href {\doibase
  10.1093/mnras/281.2.531} {\bibfield  {journal} {\bibinfo  {journal} {Mon.
  Not. Roy. Astron. Soc.}\ }\textbf {\bibinfo {volume} {281}},\ \bibinfo
  {pages} {531} (\bibinfo {year} {1996})},\ \Eprint
  {http://arxiv.org/abs/astro-ph/9511057} {arXiv:astro-ph/9511057 [astro-ph]}
  \BibitemShut {NoStop}%
\bibitem [{\citenamefont {Le~Delliou}(2006)}]{LeDelliou:2005ig}%
  \BibitemOpen
  \bibfield  {author} {\bibinfo {author} {\bibfnamefont {M.}~\bibnamefont
  {Le~Delliou}},\ }\href {\doibase 10.1088/1475-7516/2006/01/021} {\bibfield
  {journal} {\bibinfo  {journal} {JCAP}\ }\textbf {\bibinfo {volume} {0601}},\
  \bibinfo {pages} {021} (\bibinfo {year} {2006})},\ \Eprint
  {http://arxiv.org/abs/astro-ph/0506200} {arXiv:astro-ph/0506200 [astro-ph]}
  \BibitemShut {NoStop}%
\bibitem [{\citenamefont {Nunes}\ and\ \citenamefont
  {Mota}(2006)}]{Nunes:2004wn}%
  \BibitemOpen
  \bibfield  {author} {\bibinfo {author} {\bibfnamefont {N.~J.}\ \bibnamefont
  {Nunes}}\ and\ \bibinfo {author} {\bibfnamefont {D.~F.}\ \bibnamefont
  {Mota}},\ }\href {\doibase 10.1111/j.1365-2966.2006.10166.x} {\bibfield
  {journal} {\bibinfo  {journal} {Mon. Not. Roy. Astron. Soc.}\ }\textbf
  {\bibinfo {volume} {368}},\ \bibinfo {pages} {751} (\bibinfo {year}
  {2006})},\ \Eprint {http://arxiv.org/abs/astro-ph/0409481}
  {arXiv:astro-ph/0409481 [astro-ph]} \BibitemShut {NoStop}%
\bibitem [{\citenamefont {{Birkhoff}}\ and\ \citenamefont
  {{Langer}}(1923)}]{1923rmpbookB}%
  \BibitemOpen
  \bibfield  {author} {\bibinfo {author} {\bibfnamefont {G.~D.}\ \bibnamefont
  {{Birkhoff}}}\ and\ \bibinfo {author} {\bibfnamefont {R.~E.}\ \bibnamefont
  {{Langer}}},\ }\href@noop {} {\emph {\bibinfo {title} {{Relativity and modern
  physics}}}}\ (\bibinfo  {publisher} {Harvard University Press},\ \bibinfo
  {year} {1923})\BibitemShut {NoStop}%
\bibitem [{\citenamefont {Savastano}\ \emph {et~al.}(2019)\citenamefont
  {Savastano}, \citenamefont {Amendola}, \citenamefont {Rubio},\ and\
  \citenamefont {Wetterich}}]{Savastano:2019zpr}%
  \BibitemOpen
  \bibfield  {author} {\bibinfo {author} {\bibfnamefont {S.}~\bibnamefont
  {Savastano}}, \bibinfo {author} {\bibfnamefont {L.}~\bibnamefont {Amendola}},
  \bibinfo {author} {\bibfnamefont {J.}~\bibnamefont {Rubio}}, \ and\ \bibinfo
  {author} {\bibfnamefont {C.}~\bibnamefont {Wetterich}},\ }\href@noop {} {\
  (\bibinfo {year} {2019})},\ \Eprint {http://arxiv.org/abs/1906.05300}
  {arXiv:1906.05300 [astro-ph.CO]} \BibitemShut {NoStop}%
\bibitem [{\citenamefont {Amendola}\ \emph {et~al.}(2013)\citenamefont
  {Amendola} \emph {et~al.}}]{Amendola:2012ys}%
  \BibitemOpen
  \bibfield  {author} {\bibinfo {author} {\bibfnamefont {L.}~\bibnamefont
  {Amendola}} \emph {et~al.} (\bibinfo {collaboration} {Euclid Theory Working
  Group}),\ }\href {\doibase 10.12942/lrr-2013-6} {\bibfield  {journal}
  {\bibinfo  {journal} {Living Rev. Rel.}\ }\textbf {\bibinfo {volume} {16}},\
  \bibinfo {pages} {6} (\bibinfo {year} {2013})},\ \Eprint
  {http://arxiv.org/abs/1206.1225} {arXiv:1206.1225 [astro-ph.CO]} \BibitemShut
  {NoStop}%
\bibitem [{\citenamefont {Pettorino}\ and\ \citenamefont
  {Baccigalupi}(2008)}]{Pettorino:2008ez}%
  \BibitemOpen
  \bibfield  {author} {\bibinfo {author} {\bibfnamefont {V.}~\bibnamefont
  {Pettorino}}\ and\ \bibinfo {author} {\bibfnamefont {C.}~\bibnamefont
  {Baccigalupi}},\ }\href {\doibase 10.1103/PhysRevD.77.103003} {\bibfield
  {journal} {\bibinfo  {journal} {Phys. Rev.}\ }\textbf {\bibinfo {volume}
  {D77}},\ \bibinfo {pages} {103003} (\bibinfo {year} {2008})},\ \Eprint
  {http://arxiv.org/abs/0802.1086} {arXiv:0802.1086 [astro-ph]} \BibitemShut
  {NoStop}%
\bibitem [{\citenamefont {Leithes}\ \emph {et~al.}(2017)\citenamefont
  {Leithes}, \citenamefont {Malik}, \citenamefont {Mulryne},\ and\
  \citenamefont {Nunes}}]{Leithes:2016xyh}%
  \BibitemOpen
  \bibfield  {author} {\bibinfo {author} {\bibfnamefont {A.}~\bibnamefont
  {Leithes}}, \bibinfo {author} {\bibfnamefont {K.~A.}\ \bibnamefont {Malik}},
  \bibinfo {author} {\bibfnamefont {D.~J.}\ \bibnamefont {Mulryne}}, \ and\
  \bibinfo {author} {\bibfnamefont {N.~J.}\ \bibnamefont {Nunes}},\ }\href
  {\doibase 10.1103/PhysRevD.95.123519} {\bibfield  {journal} {\bibinfo
  {journal} {Phys. Rev.}\ }\textbf {\bibinfo {volume} {D95}},\ \bibinfo {pages}
  {123519} (\bibinfo {year} {2017})},\ \Eprint
  {http://arxiv.org/abs/1608.00908} {arXiv:1608.00908 [astro-ph.CO]}
  \BibitemShut {NoStop}%
\bibitem [{\citenamefont {Sapa}\ \emph {et~al.}(2018)\citenamefont {Sapa},
  \citenamefont {Karwan},\ and\ \citenamefont {Mota}}]{Sapa:2018jja}%
  \BibitemOpen
  \bibfield  {author} {\bibinfo {author} {\bibfnamefont {S.}~\bibnamefont
  {Sapa}}, \bibinfo {author} {\bibfnamefont {K.}~\bibnamefont {Karwan}}, \ and\
  \bibinfo {author} {\bibfnamefont {D.~F.}\ \bibnamefont {Mota}},\ }\href
  {\doibase 10.1103/PhysRevD.98.023528} {\bibfield  {journal} {\bibinfo
  {journal} {Phys. Rev.}\ }\textbf {\bibinfo {volume} {D98}},\ \bibinfo {pages}
  {023528} (\bibinfo {year} {2018})},\ \Eprint
  {http://arxiv.org/abs/1803.02299} {arXiv:1803.02299 [astro-ph.CO]}
  \BibitemShut {NoStop}%
\bibitem [{\citenamefont {Bekenstein}(1993)}]{Bekenstein:1992pj}%
  \BibitemOpen
  \bibfield  {author} {\bibinfo {author} {\bibfnamefont {J.~D.}\ \bibnamefont
  {Bekenstein}},\ }\href {\doibase 10.1103/PhysRevD.48.3641} {\bibfield
  {journal} {\bibinfo  {journal} {Phys. Rev.}\ }\textbf {\bibinfo {volume}
  {D48}},\ \bibinfo {pages} {3641} (\bibinfo {year} {1993})},\ \Eprint
  {http://arxiv.org/abs/gr-qc/9211017} {arXiv:gr-qc/9211017 [gr-qc]}
  \BibitemShut {NoStop}%
\bibitem [{\citenamefont {Zumalacárregui}\ and\ \citenamefont
  {García-Bellido}(2014)}]{Zumalacarregui:2013pma}%
  \BibitemOpen
  \bibfield  {author} {\bibinfo {author} {\bibfnamefont {M.}~\bibnamefont
  {Zumalacárregui}}\ and\ \bibinfo {author} {\bibfnamefont {J.}~\bibnamefont
  {García-Bellido}},\ }\href {\doibase 10.1103/PhysRevD.89.064046} {\bibfield
  {journal} {\bibinfo  {journal} {Phys. Rev.}\ }\textbf {\bibinfo {volume}
  {D89}},\ \bibinfo {pages} {064046} (\bibinfo {year} {2014})},\ \Eprint
  {http://arxiv.org/abs/1308.4685} {arXiv:1308.4685 [gr-qc]} \BibitemShut
  {NoStop}%
\bibitem [{\citenamefont {Press}\ and\ \citenamefont
  {Schechter}(1974)}]{Press:1973iz}%
  \BibitemOpen
  \bibfield  {author} {\bibinfo {author} {\bibfnamefont {W.~H.}\ \bibnamefont
  {Press}}\ and\ \bibinfo {author} {\bibfnamefont {P.}~\bibnamefont
  {Schechter}},\ }\href {\doibase 10.1086/152650} {\bibfield  {journal}
  {\bibinfo  {journal} {Astrophys. J.}\ }\textbf {\bibinfo {volume} {187}},\
  \bibinfo {pages} {425} (\bibinfo {year} {1974})}\BibitemShut {NoStop}%
\bibitem [{\citenamefont {Viana}\ and\ \citenamefont
  {Liddle}(1996)}]{Viana:1995yv}%
  \BibitemOpen
  \bibfield  {author} {\bibinfo {author} {\bibfnamefont {P.~T.~P.}\
  \bibnamefont {Viana}}\ and\ \bibinfo {author} {\bibfnamefont {A.~R.}\
  \bibnamefont {Liddle}},\ }\href {\doibase 10.1093/mnras/281.1.323} {\bibfield
   {journal} {\bibinfo  {journal} {Mon. Not. Roy. Astron. Soc.}\ }\textbf
  {\bibinfo {volume} {281}},\ \bibinfo {pages} {323} (\bibinfo {year}
  {1996})},\ \Eprint {http://arxiv.org/abs/astro-ph/9511007}
  {arXiv:astro-ph/9511007 [astro-ph]} \BibitemShut {NoStop}%
\bibitem [{\citenamefont {Manera}\ and\ \citenamefont
  {Mota}(2006)}]{Manera:2005ct}%
  \BibitemOpen
  \bibfield  {author} {\bibinfo {author} {\bibfnamefont {M.}~\bibnamefont
  {Manera}}\ and\ \bibinfo {author} {\bibfnamefont {D.~F.}\ \bibnamefont
  {Mota}},\ }\href {\doibase 10.1111/j.1365-2966.2006.10774.x} {\bibfield
  {journal} {\bibinfo  {journal} {Mon. Not. Roy. Astron. Soc.}\ }\textbf
  {\bibinfo {volume} {371}},\ \bibinfo {pages} {1373} (\bibinfo {year}
  {2006})},\ \Eprint {http://arxiv.org/abs/astro-ph/0504519}
  {arXiv:astro-ph/0504519 [astro-ph]} \BibitemShut {NoStop}%
\bibitem [{\citenamefont {Tarrant}\ \emph {et~al.}(2012)\citenamefont
  {Tarrant}, \citenamefont {van~de Bruck}, \citenamefont {Copeland},\ and\
  \citenamefont {Green}}]{PhysRevD.85.023503}%
  \BibitemOpen
  \bibfield  {author} {\bibinfo {author} {\bibfnamefont {E.~R.~M.}\
  \bibnamefont {Tarrant}}, \bibinfo {author} {\bibfnamefont {C.}~\bibnamefont
  {van~de Bruck}}, \bibinfo {author} {\bibfnamefont {E.~J.}\ \bibnamefont
  {Copeland}}, \ and\ \bibinfo {author} {\bibfnamefont {A.~M.}\ \bibnamefont
  {Green}},\ }\href {\doibase 10.1103/PhysRevD.85.023503} {\bibfield  {journal}
  {\bibinfo  {journal} {Phys. Rev. D}\ }\textbf {\bibinfo {volume} {85}},\
  \bibinfo {pages} {023503} (\bibinfo {year} {2012})}\BibitemShut {NoStop}%
\bibitem [{\citenamefont {Pace}\ \emph
  {et~al.}(2010{\natexlab{b}})\citenamefont {Pace}, \citenamefont {Waizmann},\
  and\ \citenamefont {Bartelmann}}]{Pace:2010sn}%
  \BibitemOpen
  \bibfield  {author} {\bibinfo {author} {\bibfnamefont {F.}~\bibnamefont
  {Pace}}, \bibinfo {author} {\bibfnamefont {J.~C.}\ \bibnamefont {Waizmann}},
  \ and\ \bibinfo {author} {\bibfnamefont {M.}~\bibnamefont {Bartelmann}},\
  }\href {\doibase 10.1111/j.1365-2966.2010.16841.x} {\bibfield  {journal}
  {\bibinfo  {journal} {Mon. Not. Roy. Astron. Soc.}\ }\textbf {\bibinfo
  {volume} {406}},\ \bibinfo {pages} {1865} (\bibinfo {year}
  {2010}{\natexlab{b}})},\ \Eprint {http://arxiv.org/abs/1005.0233}
  {arXiv:1005.0233 [astro-ph.CO]} \BibitemShut {NoStop}%
\bibitem [{\citenamefont {Devi}\ \emph {et~al.}(2013)\citenamefont {Devi},
  \citenamefont {Choudhury},\ and\ \citenamefont {Sen}}]{Devi:2011gb}%
  \BibitemOpen
  \bibfield  {author} {\bibinfo {author} {\bibfnamefont {N.~C.}\ \bibnamefont
  {Devi}}, \bibinfo {author} {\bibfnamefont {T.~R.}\ \bibnamefont {Choudhury}},
  \ and\ \bibinfo {author} {\bibfnamefont {A.~A.}\ \bibnamefont {Sen}},\ }\href
  {\doibase 10.1093/mnras/stt570} {\bibfield  {journal} {\bibinfo  {journal}
  {Mon. Not. Roy. Astron. Soc.}\ }\textbf {\bibinfo {volume} {432}},\ \bibinfo
  {pages} {1513} (\bibinfo {year} {2013})},\ \Eprint
  {http://arxiv.org/abs/1112.0728} {arXiv:1112.0728 [astro-ph.CO]} \BibitemShut
  {NoStop}%
\bibitem [{\citenamefont {Sheth}\ and\ \citenamefont
  {Tormen}(1999)}]{Sheth:1999mn}%
  \BibitemOpen
  \bibfield  {author} {\bibinfo {author} {\bibfnamefont {R.~K.}\ \bibnamefont
  {Sheth}}\ and\ \bibinfo {author} {\bibfnamefont {G.}~\bibnamefont {Tormen}},\
  }\href {\doibase 10.1046/j.1365-8711.1999.02692.x} {\bibfield  {journal}
  {\bibinfo  {journal} {Mon. Not. Roy. Astron. Soc.}\ }\textbf {\bibinfo
  {volume} {308}},\ \bibinfo {pages} {119} (\bibinfo {year} {1999})},\ \Eprint
  {http://arxiv.org/abs/astro-ph/9901122} {arXiv:astro-ph/9901122 [astro-ph]}
  \BibitemShut {NoStop}%
\bibitem [{\citenamefont {Reed}\ \emph {et~al.}(2007)\citenamefont {Reed},
  \citenamefont {Bower}, \citenamefont {Frenk}, \citenamefont {Jenkins},\ and\
  \citenamefont {Theuns}}]{Reed:2006rw}%
  \BibitemOpen
  \bibfield  {author} {\bibinfo {author} {\bibfnamefont {D.}~\bibnamefont
  {Reed}}, \bibinfo {author} {\bibfnamefont {R.}~\bibnamefont {Bower}},
  \bibinfo {author} {\bibfnamefont {C.}~\bibnamefont {Frenk}}, \bibinfo
  {author} {\bibfnamefont {A.}~\bibnamefont {Jenkins}}, \ and\ \bibinfo
  {author} {\bibfnamefont {T.}~\bibnamefont {Theuns}},\ }\href {\doibase
  10.1111/j.1365-2966.2006.11204.x} {\bibfield  {journal} {\bibinfo  {journal}
  {Mon. Not. Roy. Astron. Soc.}\ }\textbf {\bibinfo {volume} {374}},\ \bibinfo
  {pages} {2} (\bibinfo {year} {2007})},\ \Eprint
  {http://arxiv.org/abs/astro-ph/0607150} {arXiv:astro-ph/0607150 [astro-ph]}
  \BibitemShut {NoStop}%
\bibitem [{\citenamefont {Devi}\ \emph {et~al.}(2014)\citenamefont {Devi},
  \citenamefont {Gonzalez},\ and\ \citenamefont {Alcaniz}}]{Devi:2014rva}%
  \BibitemOpen
  \bibfield  {author} {\bibinfo {author} {\bibfnamefont {N.~C.}\ \bibnamefont
  {Devi}}, \bibinfo {author} {\bibfnamefont {J.~E.}\ \bibnamefont {Gonzalez}},
  \ and\ \bibinfo {author} {\bibfnamefont {J.~S.}\ \bibnamefont {Alcaniz}},\
  }\href {\doibase 10.1088/1475-7516/2014/06/055} {\bibfield  {journal}
  {\bibinfo  {journal} {JCAP}\ }\textbf {\bibinfo {volume} {1406}},\ \bibinfo
  {pages} {055} (\bibinfo {year} {2014})},\ \Eprint
  {http://arxiv.org/abs/1402.2590} {arXiv:1402.2590 [astro-ph.CO]} \BibitemShut
  {NoStop}%
\bibitem [{\citenamefont {Bhattacharya}\ \emph {et~al.}(2011)\citenamefont
  {Bhattacharya}, \citenamefont {Heitmann}, \citenamefont {White},
  \citenamefont {Lukic}, \citenamefont {Wagner},\ and\ \citenamefont
  {Habib}}]{Bhattacharya:2010wy}%
  \BibitemOpen
  \bibfield  {author} {\bibinfo {author} {\bibfnamefont {S.}~\bibnamefont
  {Bhattacharya}}, \bibinfo {author} {\bibfnamefont {K.}~\bibnamefont
  {Heitmann}}, \bibinfo {author} {\bibfnamefont {M.}~\bibnamefont {White}},
  \bibinfo {author} {\bibfnamefont {Z.}~\bibnamefont {Lukic}}, \bibinfo
  {author} {\bibfnamefont {C.}~\bibnamefont {Wagner}}, \ and\ \bibinfo {author}
  {\bibfnamefont {S.}~\bibnamefont {Habib}},\ }\href {\doibase
  10.1088/0004-637X/732/2/122} {\bibfield  {journal} {\bibinfo  {journal}
  {Astrophys. J.}\ }\textbf {\bibinfo {volume} {732}},\ \bibinfo {pages} {122}
  (\bibinfo {year} {2011})},\ \Eprint {http://arxiv.org/abs/1005.2239}
  {arXiv:1005.2239 [astro-ph.CO]} \BibitemShut {NoStop}%
\bibitem [{\citenamefont {Carlstrom}\ \emph {et~al.}(2011)\citenamefont
  {Carlstrom} \emph {et~al.}}]{Carlstrom:2009um}%
  \BibitemOpen
  \bibfield  {author} {\bibinfo {author} {\bibfnamefont {J.~E.}\ \bibnamefont
  {Carlstrom}} \emph {et~al.},\ }\href {\doibase 10.1086/659879} {\bibfield
  {journal} {\bibinfo  {journal} {Publ. Astron. Soc. Pac.}\ }\textbf {\bibinfo
  {volume} {123}},\ \bibinfo {pages} {568} (\bibinfo {year} {2011})},\ \Eprint
  {http://arxiv.org/abs/0907.4445} {arXiv:0907.4445 [astro-ph.IM]} \BibitemShut
  {NoStop}%
\bibitem [{\citenamefont {Sunyaev}\ and\ \citenamefont
  {Zeldovich}(1972)}]{Sunyaev:1972eq}%
  \BibitemOpen
  \bibfield  {author} {\bibinfo {author} {\bibfnamefont {R.~A.}\ \bibnamefont
  {Sunyaev}}\ and\ \bibinfo {author} {\bibfnamefont {{\relax Ya}.~B.}\
  \bibnamefont {Zeldovich}},\ }\href@noop {} {\bibfield  {journal} {\bibinfo
  {journal} {Comments Astrophys. Space Phys.}\ }\textbf {\bibinfo {volume}
  {4}},\ \bibinfo {pages} {173} (\bibinfo {year} {1972})}\BibitemShut {NoStop}%
\bibitem [{\citenamefont {Schaffer}\ \emph {et~al.}(2011)\citenamefont
  {Schaffer} \emph {et~al.}}]{Schaffer:2011mz}%
  \BibitemOpen
  \bibfield  {author} {\bibinfo {author} {\bibfnamefont {K.~K.}\ \bibnamefont
  {Schaffer}} \emph {et~al.},\ }\href {\doibase 10.1088/0004-637X/743/1/90}
  {\bibfield  {journal} {\bibinfo  {journal} {Astrophys. J.}\ }\textbf
  {\bibinfo {volume} {743}},\ \bibinfo {pages} {90} (\bibinfo {year} {2011})},\
  \Eprint {http://arxiv.org/abs/1111.7245} {arXiv:1111.7245 [astro-ph.CO]}
  \BibitemShut {NoStop}%
\bibitem [{\citenamefont {Bocquet}\ \emph {et~al.}(2019)\citenamefont {Bocquet}
  \emph {et~al.}}]{Bocquet:2018ukq}%
  \BibitemOpen
  \bibfield  {author} {\bibinfo {author} {\bibfnamefont {S.}~\bibnamefont
  {Bocquet}} \emph {et~al.} (\bibinfo {collaboration} {SPT}),\ }\href {\doibase
  10.3847/1538-4357/ab1f10} {\bibfield  {journal} {\bibinfo  {journal}
  {Astrophys. J.}\ }\textbf {\bibinfo {volume} {878}},\ \bibinfo {pages} {55}
  (\bibinfo {year} {2019})},\ \Eprint {http://arxiv.org/abs/1812.01679}
  {arXiv:1812.01679 [astro-ph.CO]} \BibitemShut {NoStop}%
\bibitem [{\citenamefont {Vanderlinde}\ \emph {et~al.}(2010)\citenamefont
  {Vanderlinde} \emph {et~al.}}]{Vanderlinde_2010}%
  \BibitemOpen
  \bibfield  {author} {\bibinfo {author} {\bibfnamefont {K.}~\bibnamefont
  {Vanderlinde}} \emph {et~al.},\ }\href {\doibase
  10.1088/0004-637x/722/2/1180} {\bibfield  {journal} {\bibinfo  {journal} {The
  Astrophysical Journal}\ }\textbf {\bibinfo {volume} {722}},\ \bibinfo {pages}
  {1180} (\bibinfo {year} {2010})}\BibitemShut {NoStop}%
\bibitem [{\citenamefont {Story}\ \emph {et~al.}(2011)\citenamefont {Story}
  \emph {et~al.}}]{Story:2011cr}%
  \BibitemOpen
  \bibfield  {author} {\bibinfo {author} {\bibfnamefont {K.}~\bibnamefont
  {Story}} \emph {et~al.},\ }\href {\doibase 10.1088/2041-8205/735/2/L36}
  {\bibfield  {journal} {\bibinfo  {journal} {Astrophys. J.}\ }\textbf
  {\bibinfo {volume} {735}},\ \bibinfo {pages} {L36} (\bibinfo {year}
  {2011})},\ \Eprint {http://arxiv.org/abs/1102.2189} {arXiv:1102.2189
  [astro-ph.CO]} \BibitemShut {NoStop}%
\bibitem [{\citenamefont {Basilakos}\ \emph {et~al.}(2010)\citenamefont
  {Basilakos}, \citenamefont {Plionis},\ and\ \citenamefont
  {Lima}}]{Basilakos:2010fb}%
  \BibitemOpen
  \bibfield  {author} {\bibinfo {author} {\bibfnamefont {S.}~\bibnamefont
  {Basilakos}}, \bibinfo {author} {\bibfnamefont {M.}~\bibnamefont {Plionis}},
  \ and\ \bibinfo {author} {\bibfnamefont {J.~A.~S.}\ \bibnamefont {Lima}},\
  }\href {\doibase 10.1103/PhysRevD.82.083517} {\bibfield  {journal} {\bibinfo
  {journal} {Phys. Rev.}\ }\textbf {\bibinfo {volume} {D82}},\ \bibinfo {pages}
  {083517} (\bibinfo {year} {2010})},\ \Eprint {http://arxiv.org/abs/1006.3418}
  {arXiv:1006.3418 [astro-ph.CO]} \BibitemShut {NoStop}%
\bibitem [{\citenamefont {Fedeli}\ \emph
  {et~al.}(2009{\natexlab{a}})\citenamefont {Fedeli}, \citenamefont
  {Moscardini},\ and\ \citenamefont {Matarrese}}]{Fedeli:2009fj}%
  \BibitemOpen
  \bibfield  {author} {\bibinfo {author} {\bibfnamefont {C.}~\bibnamefont
  {Fedeli}}, \bibinfo {author} {\bibfnamefont {L.}~\bibnamefont {Moscardini}},
  \ and\ \bibinfo {author} {\bibfnamefont {S.}~\bibnamefont {Matarrese}},\
  }\href {\doibase 10.1111/j.1365-2966.2009.15042.x} {\bibfield  {journal}
  {\bibinfo  {journal} {Mon. Not. Roy. Astron. Soc.}\ }\textbf {\bibinfo
  {volume} {397}},\ \bibinfo {pages} {1125} (\bibinfo {year}
  {2009}{\natexlab{a}})},\ \Eprint {http://arxiv.org/abs/0904.3248}
  {arXiv:0904.3248 [astro-ph.CO]} \BibitemShut {NoStop}%
\bibitem [{\citenamefont {Fedeli}\ \emph
  {et~al.}(2009{\natexlab{b}})\citenamefont {Fedeli}, \citenamefont
  {Moscardini},\ and\ \citenamefont {Bartelmann}}]{Fedeli:2008fh}%
  \BibitemOpen
  \bibfield  {author} {\bibinfo {author} {\bibfnamefont {C.}~\bibnamefont
  {Fedeli}}, \bibinfo {author} {\bibfnamefont {L.}~\bibnamefont {Moscardini}},
  \ and\ \bibinfo {author} {\bibfnamefont {M.}~\bibnamefont {Bartelmann}},\
  }\href {\doibase 10.1051/0004-6361/200811477} {\bibfield  {journal} {\bibinfo
   {journal} {Astron. Astrophys.}\ }\textbf {\bibinfo {volume} {500}},\
  \bibinfo {pages} {667} (\bibinfo {year} {2009}{\natexlab{b}})},\ \Eprint
  {http://arxiv.org/abs/0812.1097} {arXiv:0812.1097 [astro-ph]} \BibitemShut
  {NoStop}%
\bibitem [{\citenamefont {Baldi}\ \emph {et~al.}(2010)\citenamefont {Baldi},
  \citenamefont {Pettorino}, \citenamefont {Robbers},\ and\ \citenamefont
  {Springel}}]{Baldi:2008ay}%
  \BibitemOpen
  \bibfield  {author} {\bibinfo {author} {\bibfnamefont {M.}~\bibnamefont
  {Baldi}}, \bibinfo {author} {\bibfnamefont {V.}~\bibnamefont {Pettorino}},
  \bibinfo {author} {\bibfnamefont {G.}~\bibnamefont {Robbers}}, \ and\
  \bibinfo {author} {\bibfnamefont {V.}~\bibnamefont {Springel}},\ }\href
  {\doibase 10.1111/j.1365-2966.2009.15987.x} {\bibfield  {journal} {\bibinfo
  {journal} {Mon. Not. Roy. Astron. Soc.}\ }\textbf {\bibinfo {volume} {403}},\
  \bibinfo {pages} {1684} (\bibinfo {year} {2010})},\ \Eprint
  {http://arxiv.org/abs/0812.3901} {arXiv:0812.3901 [astro-ph]} \BibitemShut
  {NoStop}%
\end{thebibliography}%

\end{document}